\documentclass[aps,prd,nofootinbib]{revtex4}
\usepackage[colorlinks=true, pdfstartview=FitV, linkcolor=blue, citecolor=red, urlcolor=magenta]{hyperref}
\usepackage{graphicx}
\usepackage{latexsym}
\usepackage{amsmath}
\usepackage{amsfonts}
\usepackage{amssymb}
\usepackage{verbatim}

\newcommand{\be}{\begin{equation}}
\newcommand{\ee}{\end{equation}}
\newcommand{\bea}{\begin{eqnarray}}
\newcommand{\eea}{\end{eqnarray}}

\newcommand{\ben}{\begin{eqnarray}}
\newcommand{\een}{\end{eqnarray}}

\begin{document}

\title{Cosmological models for $f(R,T) - \Lambda(\phi)$ gravity}


\author{$^{1}$Joao R. L. Santos}
\email{joaorafael@df.ufcg.edu.br}

\author{$^2$ S. Santos da Costa}
\email{simony.santosdacosta@pi.infn.it}

\author{$^{3}$Romario S. Santos}
\email{romariofisico@gmail.com}

\affiliation{$^{1}$ Unidade Acad\^{e}mica de F\'{\i}sica, Universidade Federal de Campina Grande,
Caixa Postal 10071, 58429-900, Campina Grande, Para\'{\i}ba, Brazil.}

\affiliation{$^{2}$ Istituto Nazionale di Fisica Nucleare (INFN) Sezione di Pisa, Largo B. Pontecorvo 3, Pisa, 56127, Italy}

\affiliation{$^{3}$Departamento de Engenharia de Processos, Universidade Federal de Campina Grande,\\
	Caixa Postal 10071, 58429-900, Campina Grande, Para\'{\i}ba, Brazil.}



\begin{abstract}
The Universe is currently in a phase of accelerated expansion, a fact that was experimentally proven in the late 1990s. Cosmological models involving scalar fields allow the description of this accelerated expansion regime in the Cosmos and present themselves as a promising alternative in the study of the inflationary eras, especially the actual one which is driven by the dark energy. In this work we use the $f(R, T) - \Lambda(\phi)$ gravity to find different cosmological scenarios for our Universe. We also introduce a new path to derive analytic cosmological models which may have a non-trivial mapping between $f$ and $T$. We show that the analytic cosmological models obtained with this approach are compatible with a good description of the radiation era. In addition, we investigated the inflationary scenario and obtained a good agreement for the scalar spectral index $n_s$. Concerning the tensor-to-scalar ratio $r$, we found promising scenarios compatible with current CMB data.  \\

 
\end{abstract}
 \maketitle


\section{Introduction}

The dark sector of our Universe is one of the main open problems in the actual science. The most recent data delivered by Planck Collaboration unveil that $69\%$ of the content of our Universe is dark energy, which is responsible for the current accelerated expansion era \cite{Planck:2018}. This data set also established that the total amount of dark matter present in our Universe is $27\%$ of its content. The standard model to describe the actual data and also the present evolution of our Universe is the $\Lambda$CDM model. However, such a model presents two fundamental problems: the huge discrepancy in the determination of the cosmological constant an the coincidence problem \cite{Weinberg:1989, Pad:2003}. In order to understand the nature of the dark sector and also to overcome the issues over the $\Lambda$CDM model, Harko et al. introduced the so-called $f(R,T)$ gravity \cite{12b}. This new theory of gravity generalizes the so-called $f(R)$ or Starobinsky gravity \cite{2b} by adding a function which can depend on the trace of the energy-momentum tensor. The function $f(T)$ can be induced by the presence of exotic imperfect fluids or quantum effects due conformal anomaly in the theory \cite{12b}. This theory of gravity has been tested in several phenomenological and theoretical approaches so far as one can see in  \cite{Sardar:2023, Pappas:2022, Bose:2022, joaof}. 

Another route to amend the issues with the $\Lambda$CDM model is through a time-dependent cosmological constant or $\Lambda(t)$, which is also known as the running vacuum model. This type of model has been studied in the literature since 1980 with the seminal paper published by Coleman et al. \cite{1c}, and more recently by Polyakov \cite{2c}, and Ranjantie et al. \cite{3c}. A running vacuum scenario can be used to explain the coincidence problem and also enables us to describe two different acceleration regimes, one for early and other for late time values \cite{Lima:1994, 6c}. Recently Santos and Moraes studied how the description of the different eras of our Universe could raise in running vacuum model driven by a scalar field - $\Lambda(\phi)$ \cite{joaolambda}. There the authors also pointed out that the continuity equation is going to be satisfied for asymptotic values of field $\phi$, since such a field imposes constant values for the Hubble parameter.

Time-dependent scalar fields have been applied to describe inflationary scenarios with great success. Since the beautiful works written by Kinney \cite{Kinney:97} and Bazeia et al. \cite{Bazeia:2006} we have seen different formulations to look for analytic inflationary models which describe the different eras our Universe passes through. Despite this success, the inflationary one scalar field models faced a dilemma after the data delivered by Planck Collaboration in 2013 \cite{Planck:2013}. As pointed by Ellis et al. \cite{Ellis:2014}, inflationary one scalar field models are not able to reproduce the actual values of the spectral index and of the tensor to scalar ratio for Einstein-Hilbert gravity. However, multi fields inflation \cite{Ellis:2014}, Lorentz violating terms \cite{Almeida:2017}, new theories of gravity \cite{Sahoo:2020} or running vacuum models \cite{joaolambda}, are techniques which allow us to rescue inflation driven by scalar fields.

In this work, we intend to couple the $f(R,T)$ with $\Lambda(\phi)$ gravity. Along with our discussions, we are going to show how these two theories of gravity can work together to describe different eras of the history of the Universe. Moreover, such a theory of gravity can recover several well-known models, such as standard General Relativity, General Relativity plus a cosmological constant, $f(R)$ gravity plus a cosmological constant, etc. Besides, we also present a new method to derive analytic cosmological parameters using an inflaton field. Such an approach recovers and generalizes the results found by Moraes and Santos in \cite{joaof}. This new method also enables us to generate analytic cosmological models for non-trivial forms of $f(T)$. Beyond this first analytic approach, we also studied the $f(R,T)$ gravity for primordial inflation, obtaining theoretical predictions for the scalar spectral index $n_s$ and the tensor-to-scalar ratio $r$. The free parameters of $n_s$ and $r$ were constrained using the last set of observable data from Planck Collaboration for the Cosmic Microwave Background. The results obtained present good agreement with Planck data, at least at $3\sigma$ C.L. for the parameter $n_s$. Therefore, the $f(R,T)$ gravity models here presented introduced a new route to rescue inflation for a one scalar field approach. 

For methodological reasons, the ideas presented in this paper are organized in the following nutshell: in section \ref{sec:01} we show the generalities of our theory of gravity, determining the Friedmann equations and the equation of motion for the inflaton field. After that, in section \ref{sec:02} we present our new method to derive analytic cosmological models. Examples of cosmological scenarios are derived and carefully analyzed in section \ref{sec:03}. In section \ref{sec:03_05} we constraint our models using observable data collected by Planck Collaboration \cite{Planck:2018}, from Cosmic Microwave Background. Such an approach unveiled the potential of $f(R,T)$ gravity to describe primordial inflation. Our final remarks and perspectives are pointed in section \ref{sec:04}.

\section{The $f(R, T)$ $-$ $\Lambda(\phi)$ gravity}
\label{sec:01}

Let us start by introducing the generalities on our theory of gravity. Such a theory combines $f(R,T)$ gravity with a cosmological constant depending on the inflaton field $\phi$, i.e. 
\begin{eqnarray}
\label{ss}
S =  \int d^4 x \sqrt{-g} \left( f(R,T) -\frac{\Lambda(\phi)}{2} + \mathcal{L} \right)\,,
\end{eqnarray}
where ${\cal L}$ is standard Lagrangian density for a real scalar field, whose specific form is
\begin{equation}
\label{sec_02:eq_01}
    {\cal L} = \frac{1}{2}\,\partial_{\mu} \phi \,\partial^{\mu} \phi-V(\phi)\,.
\end{equation}

Here $f(R,T)$ gravity can exhibit non-linear geometric terms generalizing the Einstein-Hilbert sector and embeds the non-trivial contributions coming from a function that depends on the trace of the energy-momentum tensor $T$. Such contributions refer to classical imprints due to quantum anisotropies after the end of primordial inflation. Moreover, the cosmological constant $\Lambda(\phi)$ enables us to evade the coincidence problem, as pointed out by Coleman and Luccia \cite{1c} in their seminal paper. The explicit form of $\Lambda(\phi)$ is such that
\begin{equation} \label{ss_01}
    \Lambda(\phi) = c_0 + c_2 \,H^{\,2}(\phi) + c_4\,H^{4}(\phi)\,,
\end{equation}
where $H=H(\phi)$ is the Hubble parameter, which is going to depend on the evolution of the inflaton field \cite{joaolambda}. 

In order to preserve the strong evidences which support General Relativity, we are going to work with 
\begin{eqnarray}
f(R,T) = f(R) + f(T) = -\frac{R}{4} + f(T)\,.
\end{eqnarray}
With this procedure we also intend to generalize the analytic models derived by Moraes and Santos in \cite{joaof}. So, by taking that our action is constant in respect of metric variations we find
\begin{eqnarray}
\label{G}
G_{\mu \nu} = 2 T_{\mu \nu}  - 2 f\,g_{\mu \nu} - 4 f^{\,\prime} \partial_{\mu} \phi \partial_\nu \phi + 2 \rho_\Lambda g_{\mu \nu}\,,
\end{eqnarray}
for
\begin{eqnarray}
T_{\mu \nu} = 2 \frac{\partial \mathcal{L}}{\partial g^{\mu \nu}} + \mathcal{L} g_{\mu \nu}\,; \qquad \rho_{\Lambda} = \frac{\Lambda}{2}\,,
\end{eqnarray}
where primes stand for derivatives in respect to $T$. Once we would like to search for cosmological solutions for the field equations, we work with the standard Friedmann-Robertson-Walker metric, which is written as
\begin{equation}
    d\,s^2= dt^2- a^2(t)\,d\Vec{r}^{\,2}\,,
\end{equation}
where $a(t)$ is the scale factor. So, from \eqref{G} and working with a time-dependent field $\phi = \phi(t)$, we can derive the Friedmann equations
\begin{eqnarray}
\label{f1}
\frac{3}{2} H^2 =  \left( \frac{1}{2} - 2 f^{\,\prime}\right) \dot{\phi}^2 - f + V + \rho_\Lambda\,; \qquad H = \frac{\dot{a}}{a}\,,
\end{eqnarray}
and
\begin{eqnarray}
\label{f1_b}
\dot{H} = - \left(1- 2 f^{\,\prime}\right)\, \dot{\phi}^2\,,
\end{eqnarray}
where $H$ is the so-called Hubble parameter. 

Now, by taking our action constant in respect to the variation of field $\phi$ we determine the equation of motion
\begin{eqnarray}\label{eq_move}
\left(1 - 2f^{\,\prime}\right) \,\left(\ddot{\phi} + 3 H \dot{\phi}\right) - 2 \dot{f}^{\,\prime} \dot{\phi} + \left(1 - 4 f^{\,\prime}\right) \tilde{V}_\phi = 0\,,
\end{eqnarray}
where
\begin{eqnarray}
\tilde{V} = V + \rho_\Lambda\,.
\end{eqnarray}
Here dots mean derivatives in respect to time, primes mean derivatives in respect to $T$, and $\tilde{V}_\phi=d\,\tilde{V}/d\,\phi$.

\section{New method to derive cosmological models}
\label{sec:02}

In this section we are going to introduce a new path to find analytic cosmological scenarios in $f(R,T)-\Lambda(\phi)$ gravity, generalizing the approach presented by Moraes and Santos in \cite{joaof}. The main advantage of this method is that it can naturally accommodate contributions due to the time-dependent cosmological constant, and also it can be used to build analytic models for different forms of the Hubble parameter $H$, and of $f(T)$. Once we would like to search for analytic cosmological models, let us consider the following definition
\begin{equation}
\label{met_01}
    H\equiv h(\phi)\,; \qquad \dot{H}=h_\phi \dot{\phi}\,,
\end{equation}
which can be substituted in Eq. \eqref{f1_b}, yielding to the constraint
\bea\label{fo1}
h_\phi =-\Big(1-2f^{\prime}\Big)\dot{\phi}\,.
\eea
Now, by defining that the field $\phi$ obeys the first-order differential equation
\bea\label{fo2}
\dot{\phi} =-W_\phi\,; \qquad W_{\phi}=\frac{dW(\phi)}{d\phi}\,,
\eea
we determine the following form for \eqref{fo1}
\bea\label{fo3}
h_\phi =\Big(1-2f^{\prime}\Big)W_\phi\,.
\eea
Here $W(\phi)$ is an arbitrary function of the field $\phi$ which is going to establish an specific form for the cosmological potential. Thus, by taking Eqs. \eqref{met_01}, \eqref{fo2}, and \eqref{fo3} into \eqref{f1} we yield to
\begin{equation}
\label{met_02}
    \tilde{V} = \frac{3}{2}\,H^{\,2}+\frac{W_\phi^2}{2}-h_\phi\,W_\phi+f\,.
\end{equation}

Therefore, once $T= T(\phi)$, then $f^{\prime}=\dfrac{df}{dT}=\dfrac{df/d\phi}{dT/d\phi}=\dfrac{f_\phi}{T_\phi}$, enabling us to write \eqref{fo3} as
\bea\label{fo5}
f_\phi=\dfrac{1}{2}T_\phi\left(1-\dfrac{h_\phi}{W_\phi}\right)\,. 
\eea
Moreover, if we are dealing with a perfect fluid, the trace of the energy-momentum tensor has the form
\bea\label{fo6}
T=\rho - 3p\,,
\eea
where
\bea\label{fo7}
\rho=\dfrac{1}{2}\dot{\phi}^2+V\,; \qquad	
p=\dfrac{1}{2}\dot{\phi}^2-V\,,
\eea
which means that
\bea\label{fo9}
T= - W_\phi^{2}+4V\,.
\eea  

Consequently, by taking the derivative of \eqref{fo9} in respect to $\phi$, we are able to rewrite \eqref{fo5} as
\bea\label{fo11}
f_\phi=\left(\dfrac{h_\phi}{W_\phi}-1\right)\,\left(W_\phi W_{\phi\phi}-2V_\phi\right)\,.
\eea
However, by taking the derivative in respect to $\phi$ of \eqref{met_02} we obtain
\bea\label{fo4}
f_\phi=W_{\phi\phi}h_\phi+W_\phi h_{\phi\phi}-W_\phi W_{\phi\phi}-3h\,h_\phi +V_\phi+\rho_{\Lambda\,\phi}\,.
\eea
So, from Eqs. \eqref{fo11} and \eqref{fo4} we find the following constraint for the cosmological potential
\bea\label{fo12}
V_\phi=\dfrac{\left(3hh_\phi-W_\phi h_{\phi\phi}-\rho_{\,\Lambda\,\phi}\right)W_\phi}{2h_\phi-W_\phi}\,.
\eea
Thus, by substituting Eq. \eqref{fo12} into \eqref{fo4} and \eqref{fo5} we determine that
\begin{subequations}
\bea\label{fo13}
f_\phi=(h_\phi -W_\phi)\Big(\dfrac{W_{\phi\phi}(h_\phi - W_\phi/2)-3hh_\phi+W_\phi h_{\phi\phi}+\rho_{\Lambda\,\phi}}{h_\phi - W_\phi/2}\Big)\,,
\eea
and
\bea\label{fo14}
T_\phi=-2W_\phi W_{\phi\phi}+\dfrac{4W_\phi\left(3hh_\phi-W_\phi W_{\phi\phi}-\rho_{\Lambda\,\phi}\right)}{2h_\phi -W_\phi}\,,
\eea
\end{subequations}
respectively. 

Therefore, by choosing specific forms for $h(\phi)$ and for $W_\phi$ we can find different families of $f(T)$ which analytic cosmological scenarios. The dependence between $f$ and $T$ can be derived analytically or numerically by depicting a parametric plot of Eqs. \eqref{fo13} and \eqref{fo14}. 


\section{Cosmological Models}
\label{sec:03}

\subsection{First Model - $f_\phi = \alpha \, T_\phi$}
As a first example we are going to work with the following definitions for $W_\phi$ and $h(\phi)$ are
\begin{equation}
\label{m01_01}
    W_\phi = b_1\left(\phi^2-1\right)\,; \qquad h(\phi) = \left(1-2\alpha\right)\, W\,,
\end{equation}
where the superpotential $W(\phi)$ is
\begin{eqnarray}
\label{m01_02}
W = b_1 \left(- \phi + \frac{\phi^3}{3}\right) + \frac{b_2}{(1-2\alpha)}\,.
\end{eqnarray}
Here $b_1$, $b_2$, and $b_3$ are free parameters, and this specific form of $W(\phi)$ corresponds to the well-known $\phi^4$ superpotential in classical field theory. Such superpotential is broadly applied in several subjects of investigations as one can see in \cite{Bazeia:2013} and references therein. 

\begin{figure}[!h]
	\centering
	\includegraphics[width=0.45\columnwidth]{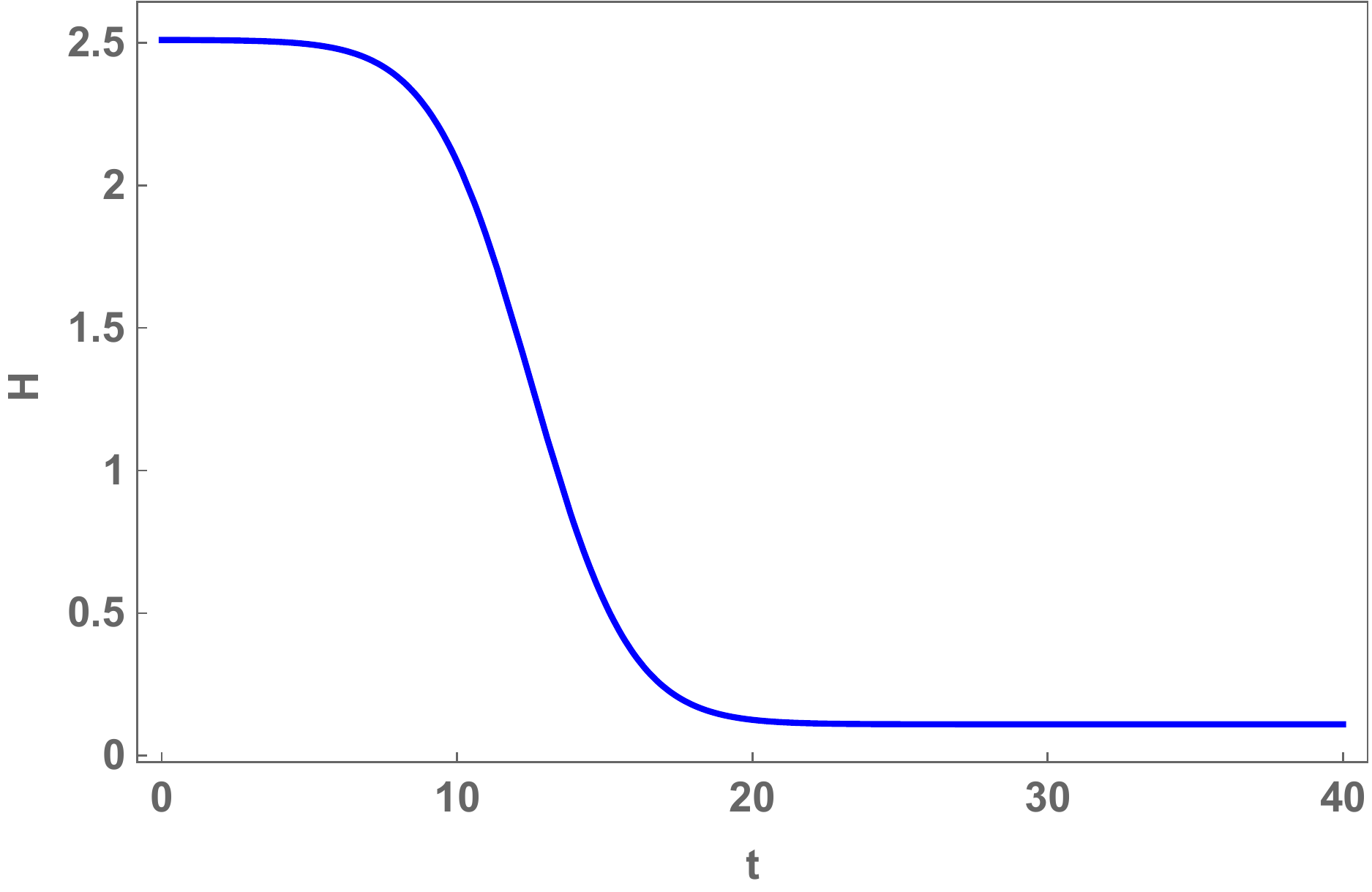}
	\caption{Time evolution of $H$ for our first model of $f(R,T) - \Lambda(t)$ gravity. The graphic was depicted for $\alpha =  -4$, $b_1=0.2$, $b_2 = 1.31$, and $b_3=-2.5$.}
	\label{H1}
\end{figure}

Thus, the correspondent first-order differential equation and its respective analytic solution are
\begin{equation}
\label{m01_03}
    \dot{\phi}=b_1\,\left(1-\phi^2\right)\,; \qquad \phi(t) = \tanh\left(b_1\,t+b_3\right)\,.
\end{equation}
Now, taking \eqref{m01_01}, and \eqref{m01_02} into Eqs. \eqref{fo12}, \eqref{fo13}, and \eqref{fo14}  we yield to
\begin{eqnarray} \label{m01_v1_01}
    &&
    V_{\phi} = -\frac{b_1 \left(\phi ^2-1\right)}{4 \alpha -1} \bigg((1-2 \alpha )^2 \,b_1\, \phi  \left(\phi ^2-3\right)+2 (2 \alpha -1) \,b_1\, \phi \\
    && \nonumber
    -2 \,c_4\, \left(\frac{1}{3} \,b_1\, \phi  \left(\phi ^2-3\right)+\frac{b_2}{1-2 \alpha }\right)^3-\frac{1}{3} \,b_1\, c_2\, \phi  \left(\phi ^2-3\right)+3 (1-2 \alpha ) \,b_2+\frac{b_2\, c_2}{2 \alpha -1}\bigg)\,,
\end{eqnarray}
\begin{eqnarray} \label{m01_01_01}
   && \nonumber
   f_{\phi} =  \frac{1}{4 \alpha -1}\bigg[4 \,\alpha \, b_1 \left(\phi ^2-1\right) \bigg(-(1-2 \alpha )^2 \,b_1\, \phi  \left(\phi ^2-3\right)+2 (1-2 \alpha ) \,b_1 \,\phi +(1-4 \alpha )\, b_1\, \phi  \\
   &&
   +2 \, c_4 \,\left(\frac{1}{3} \,b_1\, \phi  \left(\phi ^2-3\right)+\frac{b_2}{1-2 \alpha }\right)^3+\frac{1}{3} \, b_1 \, c_2 \, \phi  \left(\phi ^2-3\right)+3 (2 \alpha -1) \,b_2+\frac{b_2 \, c_2}{1-2 \alpha }\bigg)\bigg]\,,
\end{eqnarray}
\begin{eqnarray} \label{m01_01_02}
    && \nonumber
    T_{\phi} = 4 \, b_1 \, \left(\phi ^2-1\right) \bigg[-\frac{1}{4 \alpha -1}\bigg((1-2 \alpha )^2 \, b_1 \, \phi \, \left(\phi ^2-3\right)+2 (2 \alpha -1) \, b_1 \, \phi  \\
    &&
    -2 \,c_4 \, \left(\frac{1}{3} \, b_1 \,  \phi  \left(\phi ^2-3\right) +\frac{b_2}{1-2 \alpha }\right)^3-\frac{1}{3} b_1 \, c_2 \, \phi  \left(\phi ^2-3\right)+3 (1-2 \alpha ) \, b_2+\frac{b_2 \,  c_2}{2 \alpha -1}\bigg)-b_1\, \phi \bigg]\,.
\end{eqnarray}
From equations \eqref{m01_01_01}, and \eqref{m01_01_02} we are able to check that $f_\phi = \alpha\,T_\phi$, which means that $f = \alpha\,T$, corroborating with the models studied by Moraes and Santos in \cite{joaof}. Then, by integrating $V_\phi$, $f_\phi$ and $T_\phi$ in respect to field $\phi$ we determine
\begin{eqnarray}
    && \nonumber \label{m01_01_03}
    V = -\frac{1}{162 (4 \alpha -1)}\bigg[\,b_1\, \phi  \bigg(\frac{6 \,b_2\, \left(\phi ^2-3\right) \left(-108 (\alpha -1) \alpha +2\, b_1^2 c_4\, \phi ^2 \left(\phi ^2-3\right)^2+9 c_2-27\right)}{2 \alpha -1} \\
    && \nonumber
    +b_1\, \phi \, \bigg(81 \left(4 \alpha  (3 \alpha -4)-3 \phi ^2+5\right) +\phi ^2 \left(27 \left((1-2 \alpha )^2 \phi ^2+6 \alpha  (5-4 \alpha )\right)-b_1^2 \,c_4\, \left(\phi ^2-3\right)^4\right)-9 \,c_2\, \left(\phi ^2-3\right)^2\bigg) \\
    && 
  -\frac{54 \,b_1 \,b_2^2 \,c_4 \phi  \left(\phi ^2-3\right)^2}{(1-2 \alpha )^2}+\frac{108 \,b_2^3 \,c_4\, \left(\phi ^2-3\right)}{(2 \alpha -1)^3}\bigg)\bigg] - \frac{(2 \alpha -1) b_1^2+3 \,b_2^2}{8 \alpha -2}\,,
\end{eqnarray}
\begin{eqnarray} \label{m01_01_04}
     &&  \nonumber
    f = - \alpha\,b_1^2+\frac{\alpha}{81 (2 \alpha -1)^3 (4 \alpha -1)}\bigg[b_1 \, \phi  \bigg(12 (1-2 \alpha )^2 b_2 \left(\phi ^2-3\right) \left(108 (\alpha -1) \alpha -2 b_1^2 c_4 \phi ^2 \left(\phi ^2-3\right)^2-9 \,c_2+27\right) \\
    && 
    +(2 \alpha -1)^3 \,b_1\, \phi  \bigg(2 \left(\phi ^2 \left(-27 (1-2 \alpha )^2 \phi ^2+324 \alpha  (2 \alpha -3)+b_1^2 \,  c_4 \, \left(\phi ^2-3\right)^4\right)+9 \,c_2\, \left(\phi ^2-3\right)^2\right) \\
    && \nonumber
    -81 \left(8 \alpha  (3 \alpha -5)-7 \phi ^2+12\right)\bigg)+108 (2 \alpha -1) b_1 \,  b_2^2 \, c_4 \,  \phi  \left(\phi ^2-3\right)^2-216 \, b_2^3 \, c_4\, \left(\phi ^2-3\right)\bigg)\bigg]- 4\,\alpha\,\frac{(2 \alpha -1) b_1^2+3 b_2^2}{8 \alpha -2} \,,
\end{eqnarray}
\begin{eqnarray}
    &&  \nonumber \label{m01_01_05}
    T = \frac{1}{81 (2 \alpha -1)^3 (4 \alpha -1)}\bigg[b_1 \, \phi  \bigg(12 (1-2 \alpha )^2 b_2 \left(\phi ^2-3\right) \left(108 (\alpha -1) \alpha -2 b_1^2 c_4 \phi ^2 \left(\phi ^2-3\right)^2-9 \,c_2+27\right) \\
    && 
    +(2 \alpha -1)^3 \,b_1\, \phi  \bigg(2 \left(\phi ^2 \left(-27 (1-2 \alpha )^2 \phi ^2+324 \alpha  (2 \alpha -3)+b_1^2 \,  c_4 \, \left(\phi ^2-3\right)^4\right)+9 \,c_2\, \left(\phi ^2-3\right)^2\right) \\
    && \nonumber
    -81 \left(8 \alpha  (3 \alpha -5)-7 \phi ^2+12\right)\bigg)+108 (2 \alpha -1) b_1 \,  b_2^2 \, c_4 \,  \phi  \left(\phi ^2-3\right)^2-216 \, b_2^3 \, c_4\, \left(\phi ^2-3\right)\bigg)\bigg] - 4\,\frac{(2 \alpha -1) b_1^2+3 b_2^2}{8 \alpha -2} - b_1^2\,,
\end{eqnarray}
respectively. Now let us move to the determination of the cosmological parameters. We firstly start with the Hubble parameter, whose form is
\begin{equation}
    H = b_2-\frac{1}{3} (2 \alpha -1) \,b_1\, \tanh (b_1\, t+b_3) \left(\tanh ^2(b_1\, t+b_3)-3\right)\,.
\end{equation}
To find the last equation we took \eqref{met_01} together with the analytic solution presented in \eqref{m01_03}. The graphic of such a parameter is presented in Fig. \ref{H1}, where we observe the presence of two inflationary eras connected, where the Hubble parameter is approximately constant. We also realize that $H$ is not affected by the presence of the cosmological constant $\Lambda(\phi)$, since its contributions were embedded in the cosmological potential and in the trace of the energy-momentum tensor, and are going to be analyzed carefully below. 

\begin{figure}[h!]
	\centering
	\includegraphics[width=0.45\columnwidth]{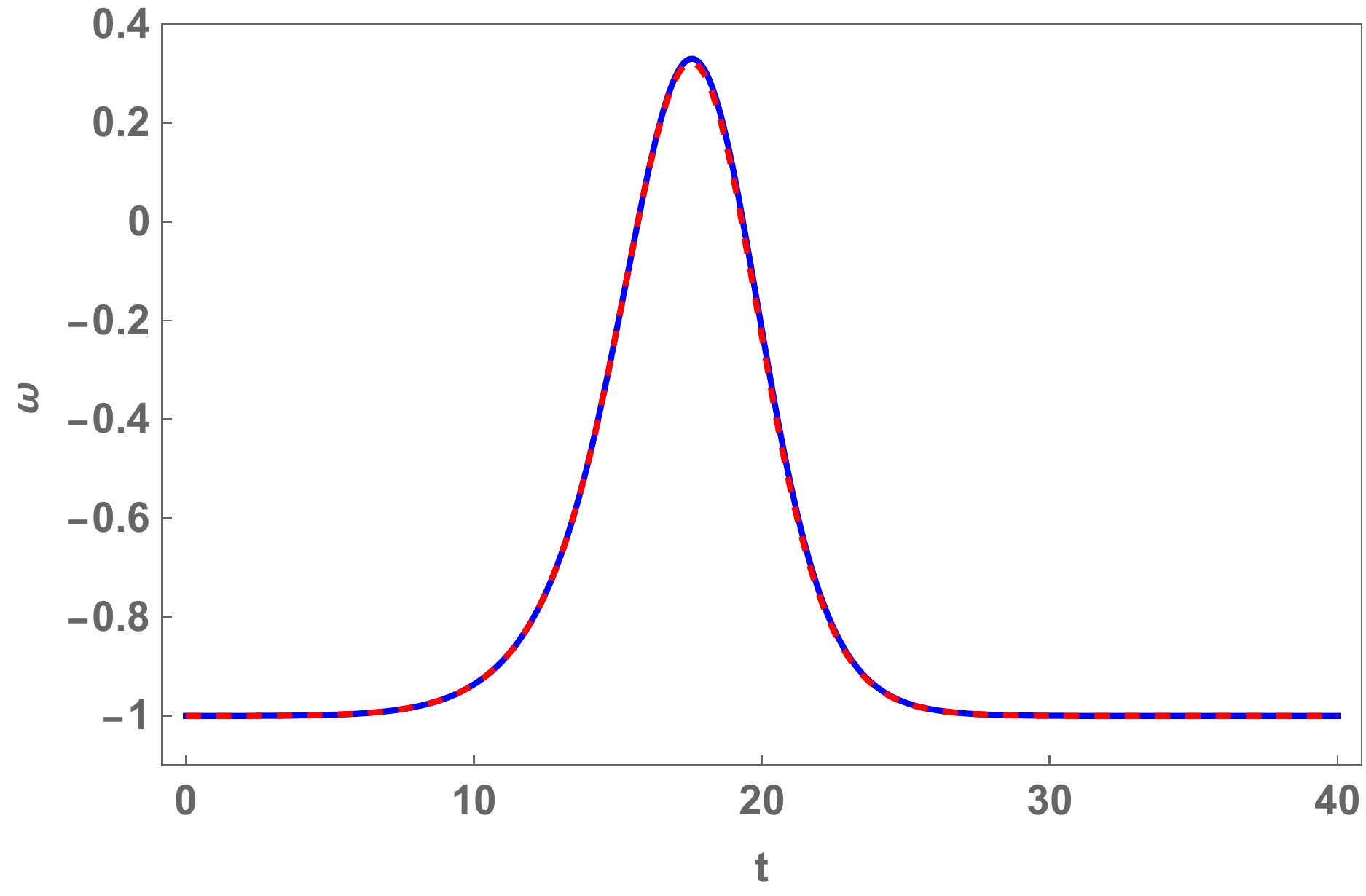}
	\caption{Equation of State parameter for our first model of $f(R,T) - \Lambda(t)$ gravity. The blue solid curve represents the case where $\Lambda = 0$, while the red dashed curve was depicted with $c_0 = 1$, $c_2 = -0.05$, and $c_4 = -0.005$. In both curves we worked with $\alpha =  -4$, $b_1=0.2$, $b_2 = 1.31$, and $b_3=-2.5$.}
	\label{w1}
\end{figure}

From the analytic form of the potential $V$, together with the analytic solution $\phi(t)$ and the definitions of the pressure and density related to the inflaton field, we can derive the Equation of State parameter $\omega$. The analytic expression of $\omega$ is going to be suppressed here for the sake of simplicity, however, its features are presented in Fig. \ref{w1}. There the dashed curve stands for a simple $f(R,T)$ model, while the blue curve represents the $f(R,T)$ gravity plus non-trivial contributions from $\Lambda(\phi)$. We can observe that the time-dependent cosmological constant can be used to make fine-tuning adjusts for $\omega$, which can be also useful to constraint the theoretical model with other cosmological parameters, such as the tensor-to-scalar ratio and the spectral index, for instance. Moreover, both forms of $\omega$ describe a transition between two different inflationary eras ($\omega \approx -1$), one for remote and the other for big values of time. We also observe that the maximum value for the EoS parameter is $\omega \approx 1/3$, corroborating with a description of the radiation era. Among these different eras, the EoS parameter presents a scenario of null pressure ($\omega = 0$), standing for the matter era. 

\begin{figure}[h!]
	\centering
	\includegraphics[width=0.45\columnwidth]{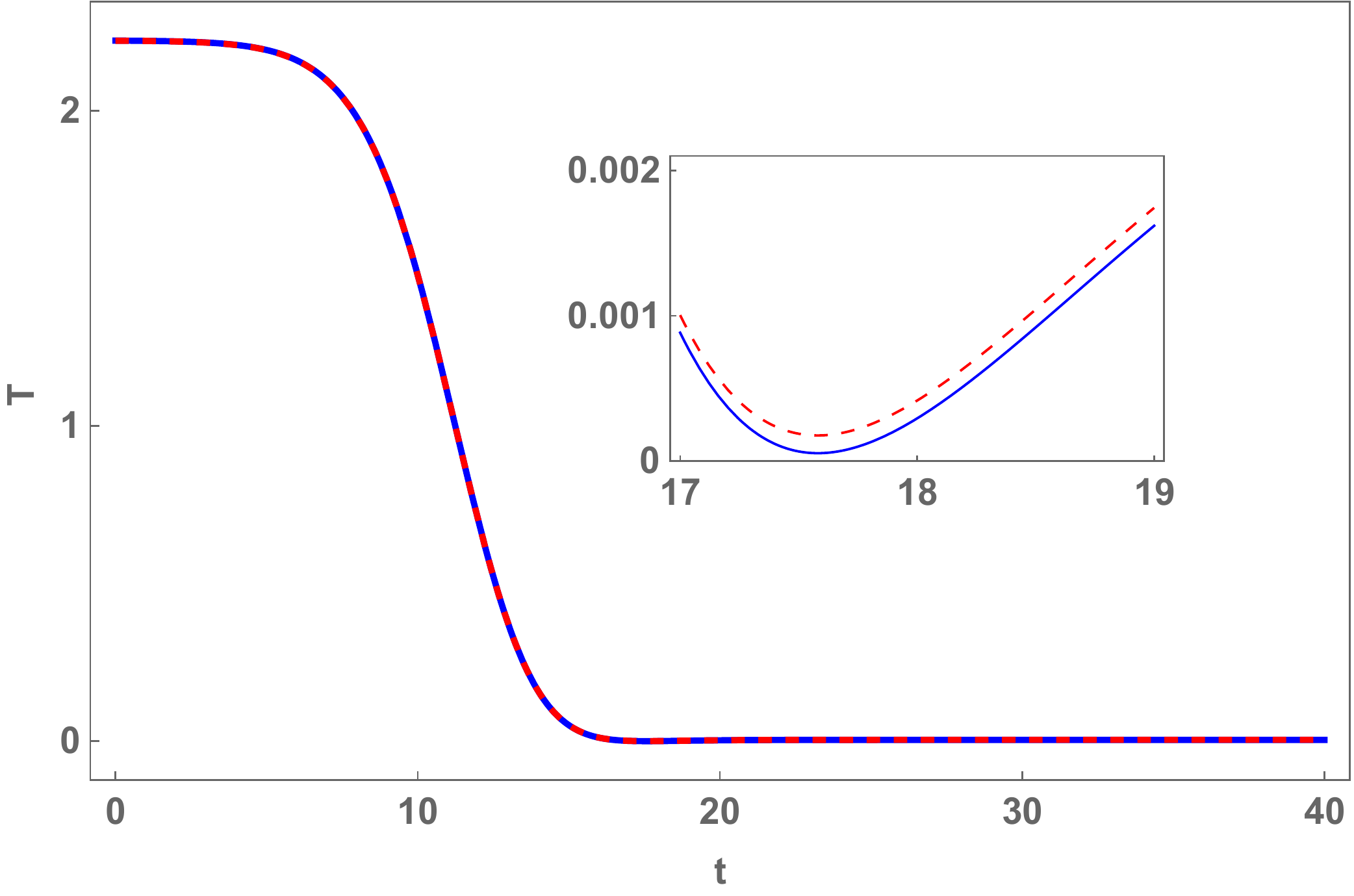}
	\caption{Trace of the energy-momentum tensor for our first model of $f(R,T) - \Lambda(t)$ gravity. The blue solid curve represents the case where $\Lambda = 0$, while the red dashed curve was depicted with $c_0 = 1$, $c_2 = -0.05$, and $c_4 = -0.005$. In both curves we worked with $\alpha =  -4$, $b_1=0.2$, $b_2 = 1.31$, and $b_3=-2.5$.}
	\label{T1}
\end{figure}

\begin{figure}[!h]
	\centering
	\includegraphics[width=0.45\columnwidth]{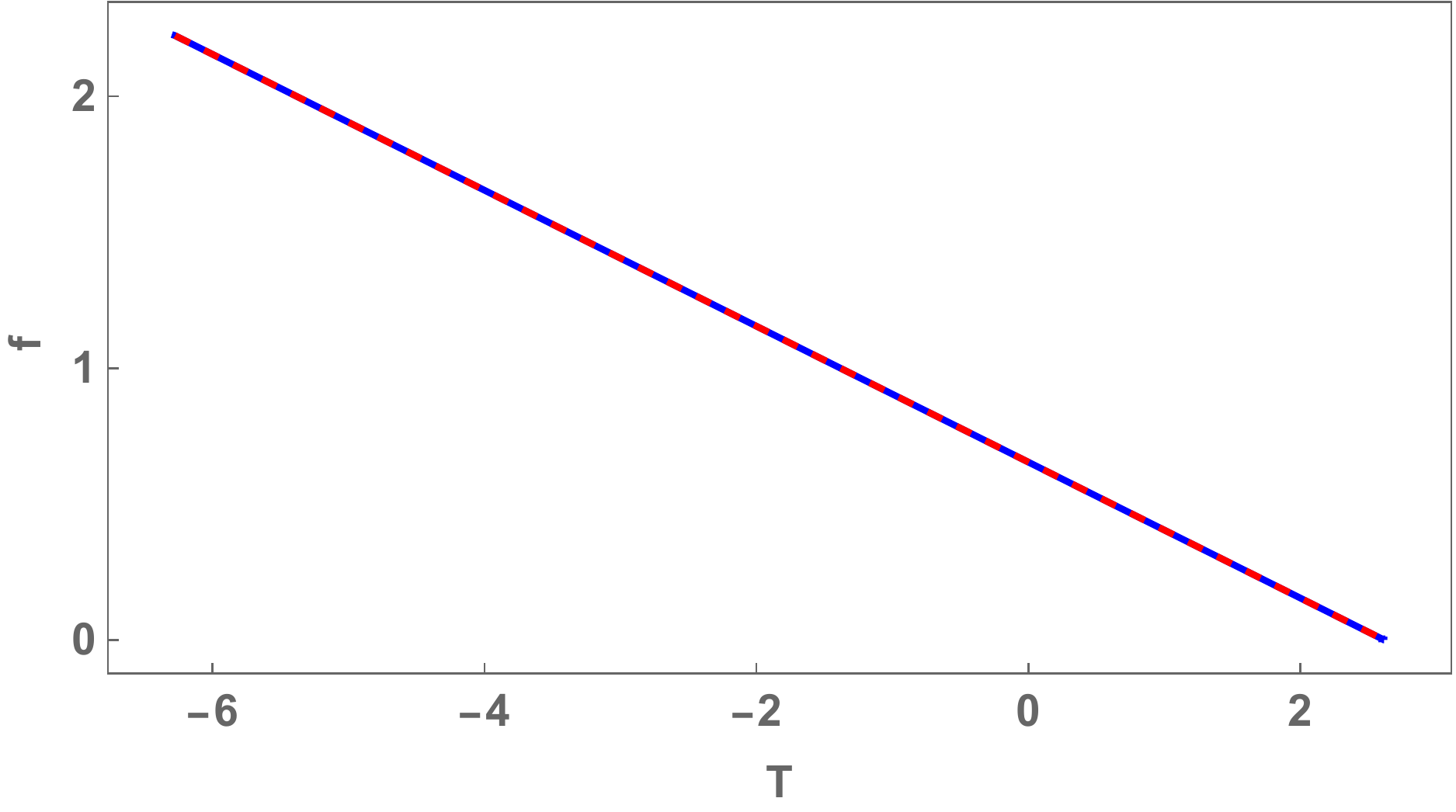}
	\caption{Parametric graph mapping $f$ and $T$ for our first model. The blue solid curve represents the case where $\Lambda = 0$, while the red dashed curve was depicted with $c_0 = 1$, $c_2 = -0.05$, and $c_4 = -0.005$. In both curves we worked with $\alpha =  -4$, $b_1=0.2$, $b_2 = 1.31$, and $b_3=-2.5$.}
	\label{parametric_01}
\end{figure}

Finally, we take \eqref{m01_01_05} and the inflaton field $\phi(t)$ to depict the behavior of the trace of the energy-momentum tensor, which is plotted in Fig. \ref{T1}. There we see that the cosmological constant $\Lambda(\phi)$ may change the minimum value of $T$ and also its asymptotic time evolution. It is relevant to point out that $T \approx 0$ when $\omega \approx 1/3$, probing the compatibility of our cosmological parameters. In order to see the functional dependence between $f$ and $T$, we build the parametric graphic presented in Fig. \ref{parametric_01}, where we used \eqref{m01_01_04}, \eqref{m01_01_05}, $\phi(t)$, and the asymptotic values of the inflaton field. This graphic confirms that $f = \alpha\, T$, corroborates with the discussions presented in \cite{joaof}.

\subsection{Second Model - $f_\phi = \left(\alpha -\beta \, b_1\, \sin (\phi )\right)\,T_{\phi}$} 

\begin{figure}[!h]
	\centering
	\includegraphics[width=0.45\columnwidth]{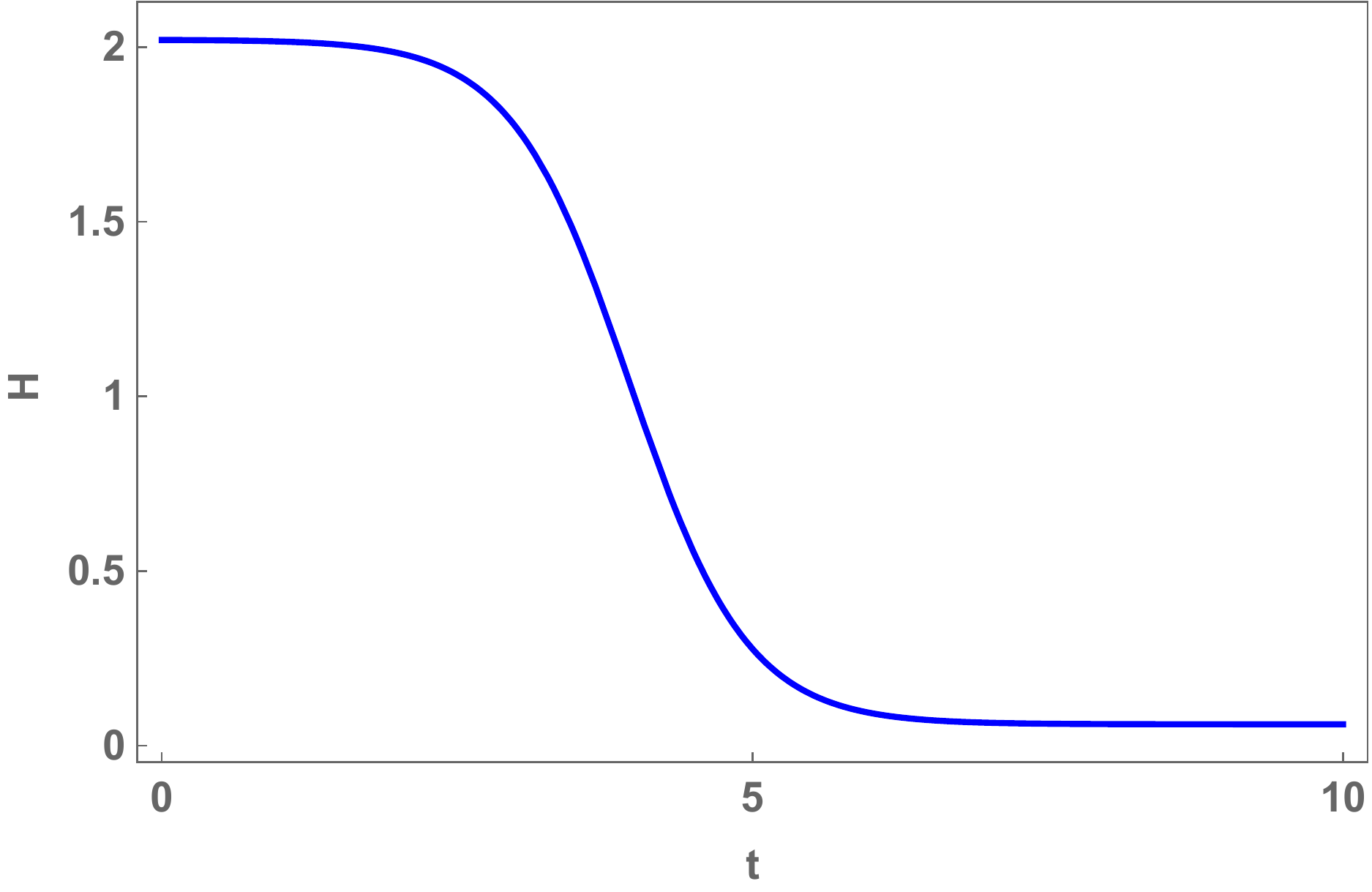}
	\caption{Time evolution of $H$ for our second model of $f(R,T) - \Lambda(t)$ gravity. The graphic was depicted for $\alpha =  0.01$, $\beta = 0.0412$, $b_1=1$, $b_2 = 1$, and $b_3=4$.}
	\label{H2}
\end{figure}

As a second example, let us present a non-trivial mapping between $f$ and $T$, resulting in a new cosmological model. In order to do so, we work with the following forms for $W(\phi)$, $W_{\phi}$, and $h(\phi)$
\begin{equation} \label{m02_01}
   W = b_1\, \sin (\phi )\,; \qquad  W_{\phi} =  b_1\, \cos (\phi )\,; \qquad h = b_2+(1-2\, \alpha )\, W +\beta \, W^2\,.
\end{equation}
This model was also investigated by Moraes and Santos in \cite{joaof}, and its known as the sine-Gordon model which has several applications in different areas of physics, mainly due to its property of integrability, as we can see in \cite{Mussardo:2004, Blas:2006, Bazeia:2011} and references therein. 
The first-order differential equation and its analytic solution for this model are given by
\begin{equation} \label{m02_02}
    \dot{\phi} = -b_1\, \cos (\phi )\,; \qquad \phi(t) = 2 \,\tan ^{-1}\left(\tanh \left(\frac{1}{2}\, (b_3-b_1\, t)\right)\right)\,.
\end{equation}
By repeating the procedure adopted in our first model, we substitute \eqref{m02_02}, and \eqref{m02_02} into Eqs. \eqref{fo12}, \eqref{fo13}, and \eqref{fo14} to determine 
\begin{eqnarray} \label{m02_03}
    &&
    V_{\phi} = \frac{b_1\,\cos(\phi)}{1-4 \alpha +4 \beta  \,b_1 \,\sin (\phi )}  \bigg(-2 \,\beta \, b_1^2 \cos ^2(\phi )+b_1\, \sin (\phi ) \left(2 \alpha  (6\, \alpha -7) \right. \\ 
    && \nonumber
   \left. +b_1\, \sin (\phi ) \left((11-18 \alpha ) \beta-2 \,b_1\, \left(c_4-3 \beta ^2\right) \sin (\phi )\right)+6 \,\beta\,  b_2-c_2+4\right)+(3-6 \alpha ) \,b_2\bigg)\,,
\end{eqnarray}
\begin{eqnarray} \label{m02_04}
    && \nonumber
    f_{\phi} = -\frac{1}{1-4 \alpha +4 \beta  b_1 \sin (\phi )} \bigg[2 b_1 \cos (\phi ) (\beta  b_1 \sin (\phi )-\alpha ) \bigg(b_1^3 \left(c_4-3 \beta ^2\right) \sin (3 \phi )+3 (6 \alpha -5) \beta  b_1^2 \cos (2 \phi ) \\
    &&
    +(11-18 \alpha ) \beta  b_1^2+ b_1 \sin (\phi ) \left(8 \alpha  (3 \alpha -4)-3 b_1^2 \left(c_4-3 \beta ^2\right)+12 \beta  b_2-2 c_2+9\right)+(6-12 \alpha ) b_2\bigg)\bigg]\,,
\end{eqnarray}
\begin{eqnarray} \label{m02_05}
    && \nonumber
    T_{\phi} = \frac{1}{1-4 \alpha +4 \beta  b_1 \sin (\phi )} \bigg[2 b_1\cos (\phi ) \bigg(-4 \beta  b_1^2 \cos ^2(\phi )+b_1 \sin (\phi ) \bigg(8 \alpha  (3 \alpha -4) \\
    &&
   +2 b_1 \sin (\phi ) \left((13-18 \alpha ) \beta -2 b_1 \left(c_4-3 \beta ^2\right) \sin (\phi )\right)+12 \beta  b_2-2 c_2+9\bigg)+6 (1-2 \alpha ) b_2\bigg)\bigg]\,.
\end{eqnarray}
After inspecting \eqref{m02_04} and \eqref{m02_05} we verify that $f_{\phi} = \left(\alpha -\beta \, b_1\, \sin (\phi )\right)\,T_{\phi}$, unveiling a non-trivial mapping between $f$ and $T$ for $\beta \neq  0$. Such mapping generalizes the classes of analytic models for $f(R,T)$ gravity. So, by integrating these previous equations in respect to $\phi$ we find
\begin{eqnarray} \label{m02_06}
    &&
    V = -\frac{1}{384 \beta ^4}\,\bigg[b_1 \bigg(\frac{3 }{b_1}\log (-4 \,\alpha +4 \,\,\beta \, b_1\, \sin (\phi )+1) \bigg(\beta ^2 \left(-16 \,\alpha ^2-32 \,\alpha +64 \,\beta ^2\, b_1^2-48 \, \beta\,  b_2\right. \\
    && \nonumber
   \left.+8 (4 \alpha -1) \,c_2+9\right) +(4 \,\alpha -1)^3 c_4\bigg)  +64 \,\beta ^3 b_1^2 \left(c_4-3 \beta ^2\right) \sin ^3(\phi )-24 \beta ^2 b_1 \sin ^2(\phi ) \left((23-24 \alpha ) \beta ^2   -4 \alpha  c_4+c_4\right)\\
    && \nonumber
 +12 \beta  \sin (\phi ) \left((1-4 \alpha )^2 c_4-\beta ^2 (4 \alpha +48 \,\beta \, b_2-8 \,c_2+9)\right)\bigg)\bigg]\,,
\end{eqnarray}
\begin{eqnarray} \label{m02_07}
    &&\nonumber
    f = \frac{-b_1}{384 \beta ^4} \bigg(\frac{3}{b_1} \log (-4 \alpha +4 \beta  b_1 \sin (\phi )+1) \left(\beta ^2 \left(-16 \alpha ^2-32 \alpha +64 \beta ^2 b_1^2-48 \beta  b_2+8 (4 \alpha -1) c_2+9\right)\right.\\
    && \nonumber
   \left. +(4 \alpha -1)^3 c_4\right)+4 \beta  \bigg(-6 \beta ^3 b_1^3 \left(c_4-3 \beta ^2\right) \cos (4 \phi )+\sin (\phi ) \left(c_4 \left(48 \alpha ^2-24 \alpha +8 \beta ^2 b_1^2+3\right) \right.\\
    && \nonumber
    \left.-3 \beta ^2 \left(-8 \beta ^2 b_1^2+4 \alpha  \left(24 \beta ^2 b_1^2+48 \beta  b_2+1\right)-48 \beta  b_2-8 c_2+9\right)\right) +\beta  b_1 \cos (2 \phi ) \left(-3 \beta ^2 \left(96 \alpha ^2-104 \alpha \right.\right.\\
    &&
    \left.\left.+24 \beta ^2 b_1^2+48 \beta  b_2-8 c_2+9\right) +3 c_4 \left(-4 \alpha +8 \beta ^2 b_1^2+1\right)-8 \beta  b_1 \sin (\phi ) \left(9 (3-4 \alpha ) \beta ^2+c_4\right)\right)\bigg)\bigg]\,,
\end{eqnarray}
\begin{eqnarray} \label{m02_08}
    &&
    T = \frac{1}{96 \beta ^4}\bigg[-96 \beta ^4 b_1^2 \cos ^2(\phi )-3 \log (-4 \alpha +4 \beta  b_1 \sin (\phi )+1) \left(\beta ^2 \left(64 \beta ^2 b_1^2-48 \beta  b_2 \right.\right. \\
    && \nonumber
    \left.\left.+(4 \alpha -1) (-4 \alpha +8 c_2-9)\right)+(4 \alpha -1)^3 c_4\right)  +4 \beta  b_1 \sin (\phi )\bigg(2 \beta  b_1 \sin (\phi ) \left(8 \beta  b_1 \left(3 \beta ^2-c_4\right) \sin (\phi ) \right. \\
    && \nonumber
   \left.+3 \left((23-24 \alpha ) \beta ^2-4 \alpha  c_4+c_4\right)\right)+3 \beta ^2 (4 \alpha +48 \beta  b_2-8 c_2+9)-3 (1-4 \alpha )^2 c_4\bigg)\bigg]\,.
\end{eqnarray}
All the previous ingredients yield us to compute our cosmological parameters. Firstly,  using $h(\phi)$ and the inflaton field $\phi(t)$ we find the Hubble parameter
\begin{equation}
    H = b_1 \sin \left(2 \tan ^{-1}\left(\tanh \left(\frac{1}{2} (b_3-b_1 t)\right)\right)\right) \left(-2 \alpha +\beta  b_1 \sin \left(2 \tan ^{-1}\left(\tanh \left(\frac{1}{2} (b_3-b_1 t)\right)\right)\right)+1\right)+b_2\,,
\end{equation}
which is presented in detail in Fig. \ref{H2}. One more time we derive a cosmological model describing a Universe which passes through two different inflationary eras, where $H \approx cte$. These results hold for both $f(R,T)$ and $f(R,T) - \Lambda(\phi)$, and also it recovers the features found by Moraes and Santos in \cite{joaof} if we take the limit $\beta \rightarrow 0$.

\begin{figure}[h!]
	\centering
	\includegraphics[width=0.45\columnwidth]{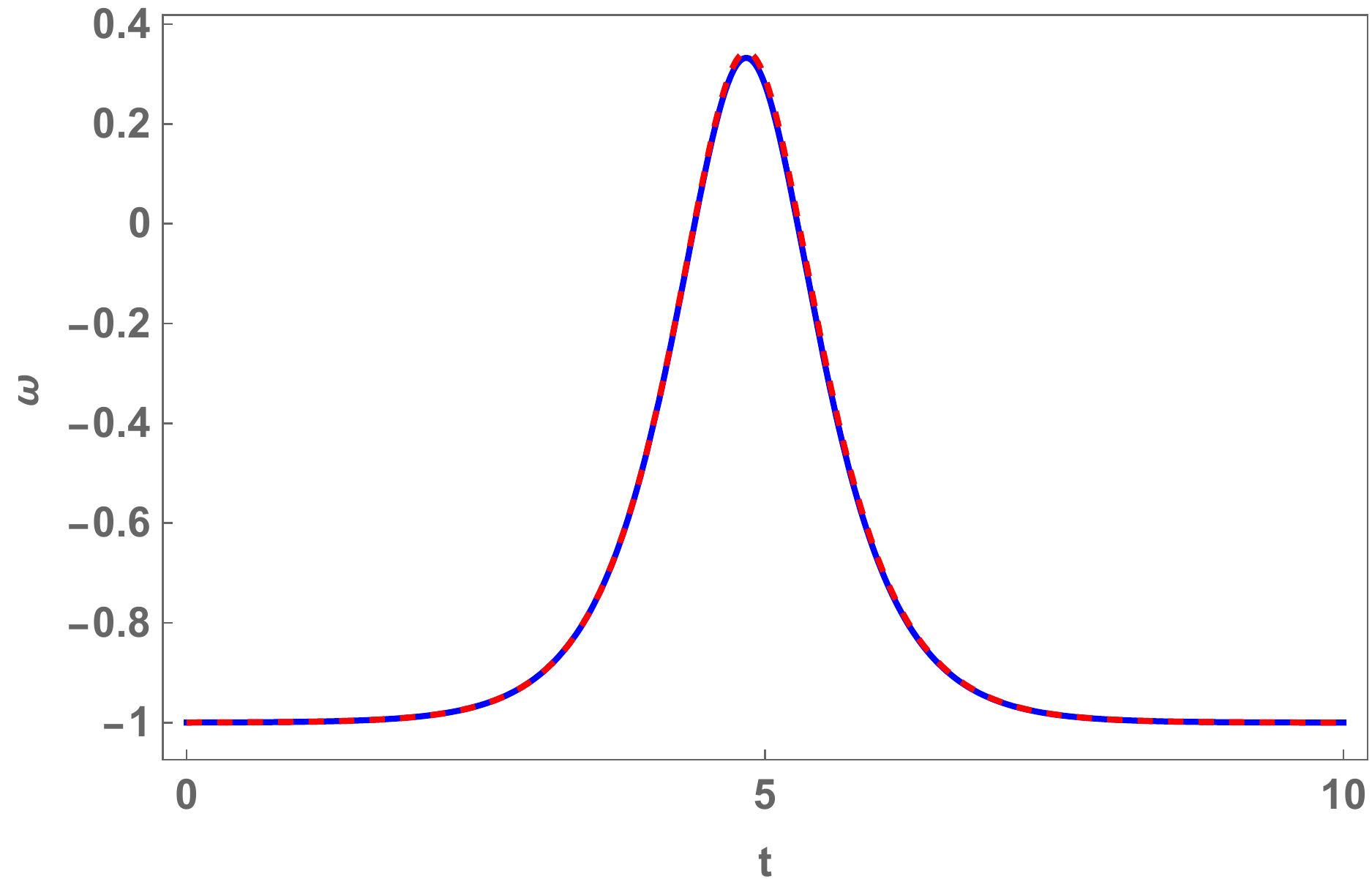}
	\caption{Equation of State parameter for our second model of $f(R,T) - \Lambda(t)$ gravity. The blue solid curve represents the case where $\Lambda = 0$, while the red dashed curve was depicted with $c_0 = 1$, $c_2 = -0.0003$, and $c_4 = -0.00003$. In both curves we worked with $\alpha =  0.01$, $\beta = 0.0412$, $b_1=1$, $b_2 = 1$, and $b_3=4$.}
	\label{w2}
\end{figure}

Using $V$ from \eqref{m02_06} together with $\phi(t)$ and the definitions for $\rho$ and $p$ we were able to derive an analytic form for the EoS parameter $\omega$, which is suppressed for the sake of simplicity. Its features are illustrated in Fig. \ref{w2}, where we can see an analogous behavior with our first model. This EoS parameter presents two inflationary eras, one for remote and another for future values of time. We also found $\omega \approx 1/3$ and $\omega =0$ which are compatible with the description of the radiation and the matter era, respectively, allowing the Universe to form complex objects such as hydrogen clouds and galaxies.

\begin{figure}[h!]
	\centering
	\includegraphics[width=0.45\columnwidth]{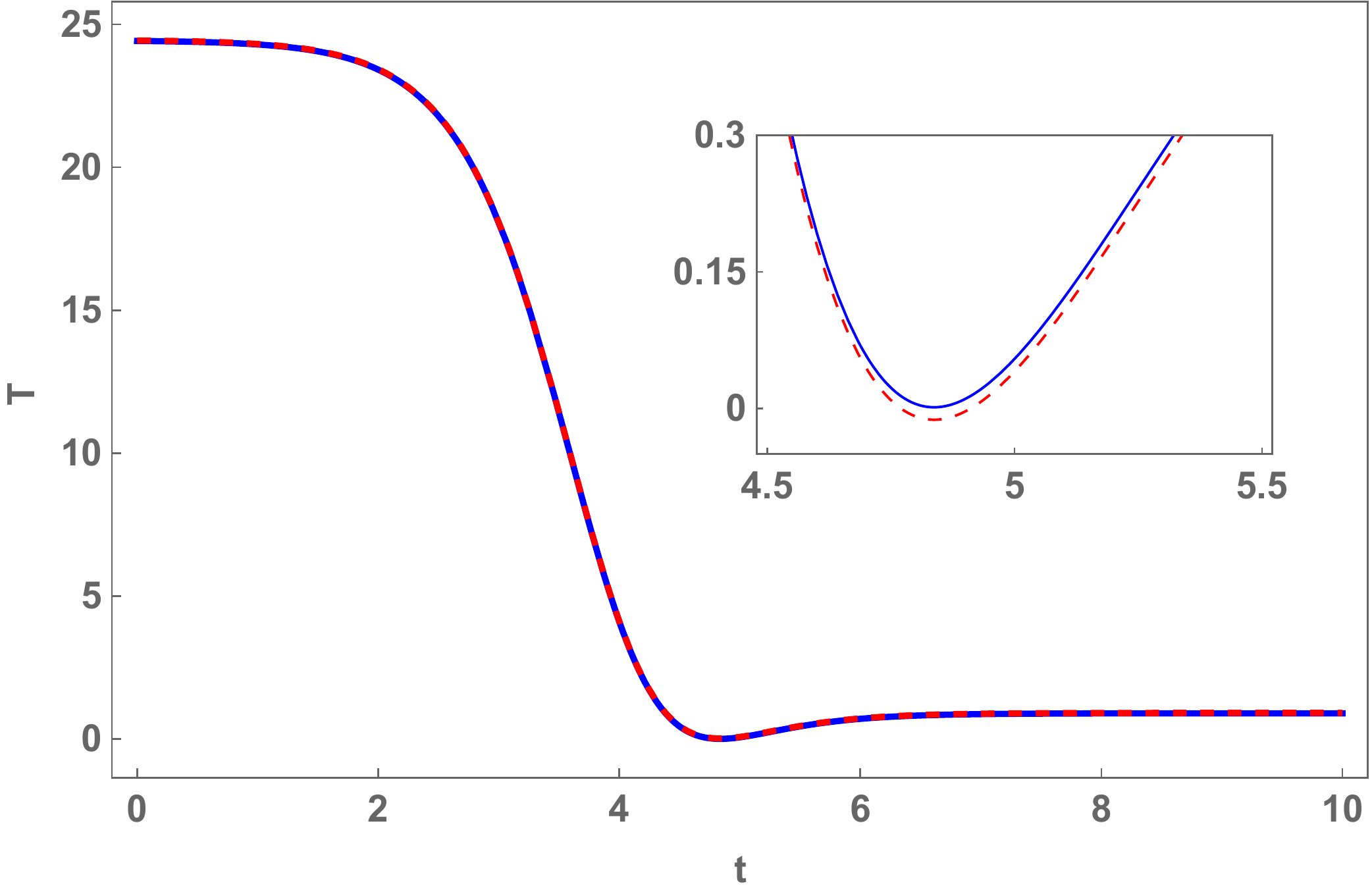}
	\caption{Trace of the energy-momentum tensor for our second model of $f(R,T) - \Lambda(t)$ gravity. The blue solid curve represents the case where $\Lambda = 0$, while the red dashed curve was depicted with $c_0 = 1$, $c_2 = -0.0003$, and $c_4 = -0.00003$. In both curves we worked with $\alpha =  0.01$, $\beta = 0.0412$, $b_1=1$, $b_2 = 1$, and $b_3=4$.}
	\label{T2}
\end{figure}

The time evolution of the trace of the energy-momentum tensor can be visualized in Fig. \ref{T2}, where we see how $\Lambda(\phi)$ can fine-tune the minimum and the asymptotic value of $T$. We also observe that $T \approx 0$ when $\omega \approx 1/3$, confirming that our cosmological parameters are describing a solid history of the evolution of the Universe.  


\begin{figure}[!h]
	\centering
	\includegraphics[width=0.45\columnwidth]{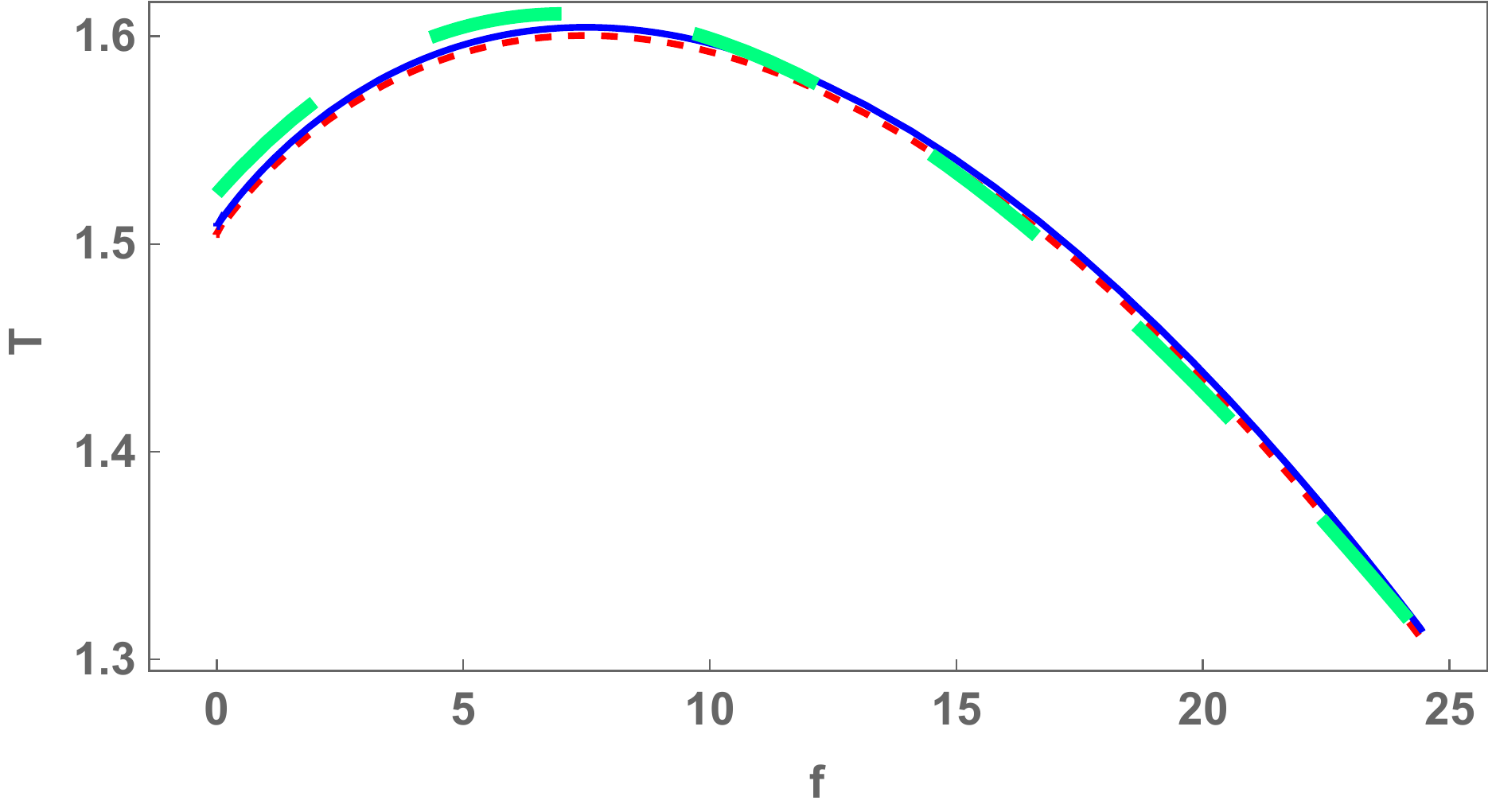} 
	\caption{Parametric plot mapping $f$ and $T$ for our second model.  The green dashed curve represents an analytic function mapping our parametric plot \eqref{m02_09}. The blue solid curve represents the case where $\Lambda = 0$, while the red dashed curve was depicted with $c_0 = 1$, $c_2 = -0.0003$, and $c_4 = -0.00003$. In both curves we worked with $\alpha =  0.01$, $\beta = 0.0412$, $b_1=1$, $b_2 = 1$, and $b_3=4$.}
	\label{parametric_02}
\end{figure}

From Eqs. \eqref{m02_07} - \eqref{m02_08} and using the analytic solution $\phi(t)$ which its asymptotic behavior we were able to derive the parametric plot shown in Fig. \ref{parametric_02}. There we are able to see the expected non-trivial functional mapping between $T$ and $f$. In order to find a functional analytic form connecting $f$ and $T$ we mapped its parametric curve with the function
\begin{equation} \label{m02_09}
    T = \alpha_4-e^{-\alpha_1 (f-\alpha_2)}-\alpha_3 (f-\alpha_2)\,,
\end{equation}
characterizing an asymmetric parabola, whose inverse function is
\begin{equation}
    f = \frac{\alpha_1 \alpha_2 \alpha_3+\alpha_1 \alpha_4+\alpha_3 {\cal W}\left(-\frac{\alpha_1 e^{\frac{\alpha_1\, T}{\alpha_3}-\frac{\alpha_1 \alpha_4}{\alpha_3}}}{\alpha_3}\right)-\alpha_1\, T}{\alpha_1 \,\alpha_3}\,,
\end{equation}
with $\alpha_1 = 0.0648 $, $\alpha_2 = 1.1352 $, $\alpha_3 = 0.0435 $, $\alpha_4 = 2.5464$, and ${\cal W}$ is the so-called Lambert or product logarithm function. This mapping is also presented in Fig. \ref{parametric_02}, where we show the optimal adjustment of the analytic function \eqref{m02_09} with our parametric curves involving $f$ and $T$ (green dashed curve).

\section{Primordial Inflation}
\label{sec:03_05}
In order to explore the viability of the new models in the context of inflationary expansion using early Universe data, let us rewrite Eq. \eqref{ss} as
\begin{equation} \label{EHact}
    S_{EH} = \int\,d^4x\,\sqrt{-g}\left(-\frac{R}{4}+\frac{1}{2}\,\partial_{\mu}\phi\,\partial^{\mu}\phi-U(\phi)\right)\,,
\end{equation}
which is the Einstein-Hilbert version of our initial action, where 
\begin{equation} \label{inf_002}
    U(\phi) = V(\phi) - f(T) +\rho_{\Lambda}\,,
\end{equation}
is the effective potential which embeds the nontrivial contributions from $f(T)$ and $\Lambda(\phi)$. Now, let us derive Eq. \eqref{inf_002} in respect to field $\phi$, yielding to
\begin{equation} \label{inf_003}
    U_{\phi} = V_{\phi} -f_{\phi} + \rho_{\Lambda\,\phi}\,,
\end{equation}
where we considered again that $f^{\prime} = f_{\phi}/T_{\phi}$. This previous procedure allows us to use Eqs. \eqref{ss_01}, \eqref{met_01}, \eqref{fo2}, \eqref{fo12}, and \eqref{fo13} to rewrite \eqref{inf_003} as
\begin{equation} \label{inf_004}
    U_\phi = W_{\phi} \left(W_{\phi\phi}-h_{\phi\phi}\right)-h_{\phi} W_{\phi\phi}+3 h\, h_{\phi}\,.
\end{equation}

Then, we can substitute Eq. \eqref{m01_01} into \eqref{inf_004}, whose integration with respect to $\phi$ leads us to the following effective potential for our first model
\begin{eqnarray}
\nonumber
    U_{1}(\phi) &=&  \frac{b_1\,\phi}{6} \bigg(b_1 \phi  \left(4 \alpha ^2 \left(\phi ^2-3\right)^2-4 \alpha  \left(\phi ^4-9 \phi ^2+15\right)+\phi ^4-9 \phi ^2+15\right) \\
    && 
    -6 \,b_2\, (2 \alpha -1) \left(\phi ^2-3\right)\bigg) -\frac{5 \left((2 \alpha -1) b_1^2+3 b_2^2\right)}{8 \alpha -2}-b_1^2+c\,.
\end{eqnarray}
Analogously, we can take Eq. \eqref{m02_01} into \eqref{inf_004}, which after one integration in respect to $\phi$ yields to
\begin{eqnarray}
    \nonumber
    U_{2}(\phi) &=& \frac{b_1}{6} \, \bigg(9 \beta ^2 b_1^3 \sin ^4(\phi )+2 (13-18 \alpha ) \beta  b_1^2 \sin ^3(\phi )-\beta  b_1^2 \sin (3 \phi ) \\
    &&
    -9 \sin (\phi ) \left(\beta  b_1^2+4 \alpha  b_2-2 b_2\right)-6 b_1 \cos ^2(\phi ) \left(6 \alpha ^2-8 \alpha +3 \beta  b_2+2\right)\bigg)+c\,,
\end{eqnarray}
as the effective potential for our second model. 

At this point, we can derive the theoretical predictions of the model studying the dynamics of the inflaton field through the Friedmann and Klein-Gordon equations\footnote{Obtained from the variation of the action \eqref{EHact} w.r.t. to the metric, $g_{\mu\nu}$, and to the field $\phi$, respectively.},
\begin{eqnarray}
    H^2 = \frac{1}{3}\left[\frac{\dot{\phi}^2}{2} + U(\phi) \right]\,,\\
    \ddot{\phi} +3H\dot{\phi} + U_\phi(\phi) = 0\,,
\end{eqnarray}
For inflation to happen we must guarantee that the expansion rate, $H$, is almost a constant that occurs when the field slowly rolls down its potential until its minimum. This is well known as the slow-roll approximation and is well translated on the slow-roll inflationary parameters, that depend on the field potential and its derivatives w.r.t the scalar field $\phi$,
\begin{eqnarray}
    \epsilon = \frac{1}{2}\frac{U_\phi(\phi)}{U(\phi)}, ~~~~~~ \mathrm{and} ~~~~~~ \eta = \frac{U_{\phi\phi}(\phi)}{U(\phi)},
\end{eqnarray}
such that inflation happens while the conditions $\epsilon, \eta \ll 1$ are satisfied, and the condition $\epsilon(\phi)=1$ defines the value of the field $\phi$ when inflation ends, $\phi_{end}$. The duration of inflation is also an important quantity - useful to estimate the inflaton field value when exists N e-folds to the end of inflation - calculated as:
\begin{eqnarray}
    N=\int_{a_{ini}}^{a_{end}}d\ln{a}=\int_{\phi_{end}}^{\phi}\frac{d\phi}{\sqrt{2\epsilon}}.
    \label{eq:Nefolds}
\end{eqnarray}
The horizon problem, for example, needs between 40 and 60 e-folds to be solved~\cite{Baumann:2018muz}. Plus, once all of these conditions are satisfied, inflation successfully happens, and we can put constraints on the model we consider. 

If we consider the CMB temperature fluctuations data, we can put constraints on both $U_1$ and $U_2$ models. To do so, look at the scalar spectral index, $n_s$, and the tensor-to-scalar ratio, $r$. The former is associated with scalar (or curvature) perturbations and can be interpreted as fluctuations in the matter density of the Universe, while the second one is linked with tensor perturbations and describes the amplitude of primordial gravitational waves. To access the physics behind the primordial perturbations contained in the fluctuations of CMB, we can use various correlation functions parameterized in terms of the perturbation power spectrum. We restrict ourselves to the primordial power spectrum of curvature perturbations, given by:
\begin{eqnarray}
    P_{\mathcal{R}}(k)=A_s\left(\frac{k}{k_*}\right)^{n_s-1} ~~~~~~ \mathrm{or} ~~~~~~ P_{\mathcal{R}}(k)=\frac{1}{12\pi^2}\frac{U^3}{M_{Pl}^6U_{\phi}^2}\bigg|_{k=k_*},
\end{eqnarray}
where $A_s$ is the scalar power spectrum amplitude and $k_*$ is a pivot scale that crosses the horizon during inflation. The scalar spectral index, $n_s$, is defined as:
\begin{eqnarray}
    n_s - 1 = \frac{d\ln{P_{\mathcal{R}}}}{d\ln{k}} =- 6\epsilon + 2\eta.
\end{eqnarray}
Moreover, the primordial power spectrum of tensor perturbations are such that 
\begin{eqnarray}
    P_{\mathcal{T}}(k)=\frac{8}{M_{Pl}^2}\left(\frac{H}{2\pi}\right)^2\left(\frac{k}{k_*}\right)^{n_T}\equiv A_T^2\left(\frac{k}{k_*}\right)^{n_T}, ~~~~~~ \mathrm{with} ~~~~~~ n_T=\frac{d\ln{P_{\mathcal{T}}}}{d\ln{k}}=-2\epsilon.
\end{eqnarray}
Note that the tensor perturbation is almost scale invariant with its amplitude depending only on the value of the Hubble parameter during inflation. This implies that it depends only on the energy scale $V^{1/4}$ associated with the inflaton potential. Therefore, the detection of gravitational waves would be a direct measure of the energy scale associated with the inflaton~\cite{rioto02}.

Finally, there is a consistency relation that is valid for inflation models driven by a single scalar field. The tensor-to-scalar ratio,
\begin{eqnarray}
    r = \frac{A_T^2}{A_s^2}=16\epsilon,
\end{eqnarray}
associated with the amplitude of primordial gravitational waves. With this whole machinery in hand, we can calculate their theoretical predictions and confront them with the CMB temperature data. 

\begin{figure}[ht!]
	\centering
	\includegraphics[width=0.45\columnwidth]{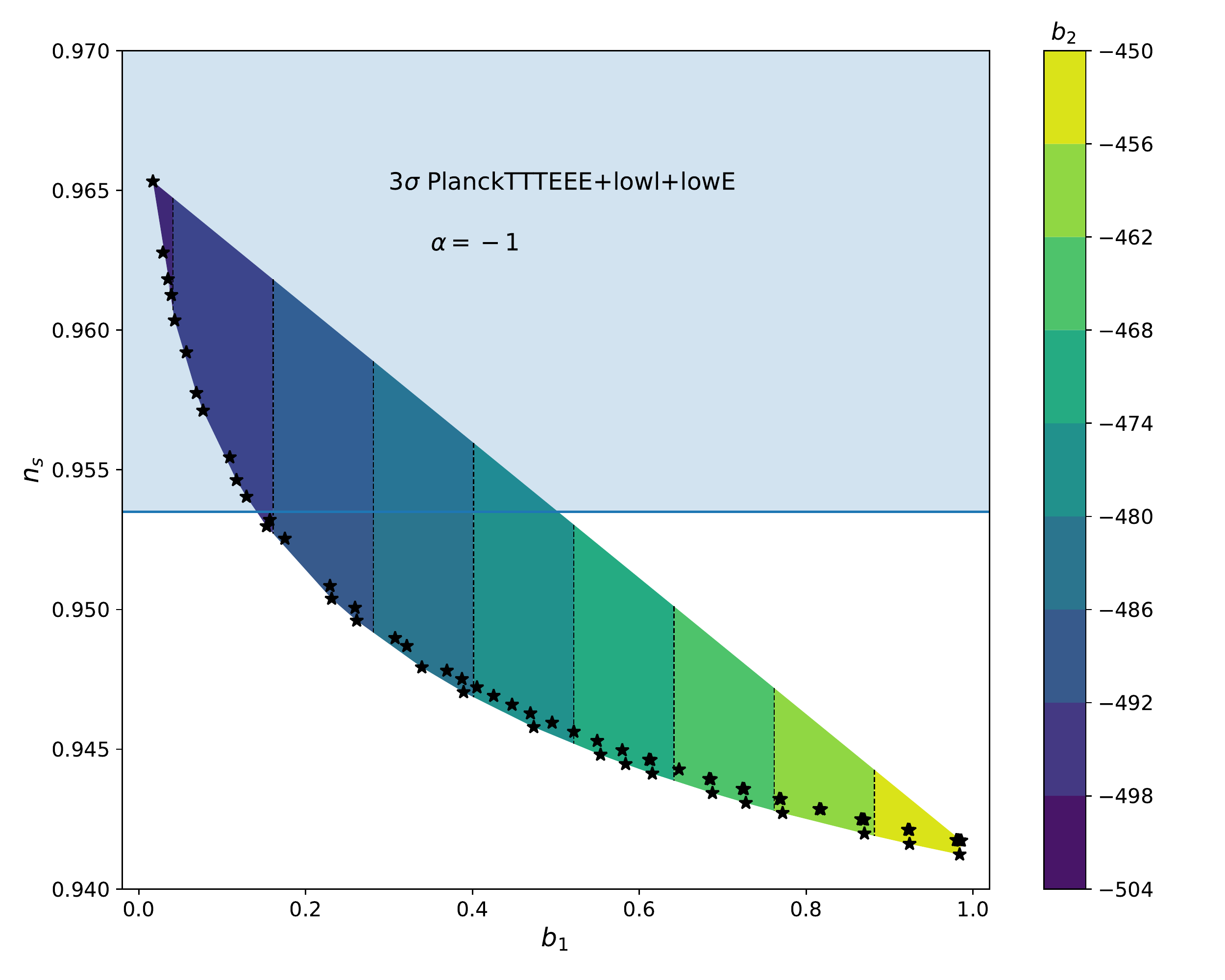}
        \includegraphics[width=0.45\columnwidth]{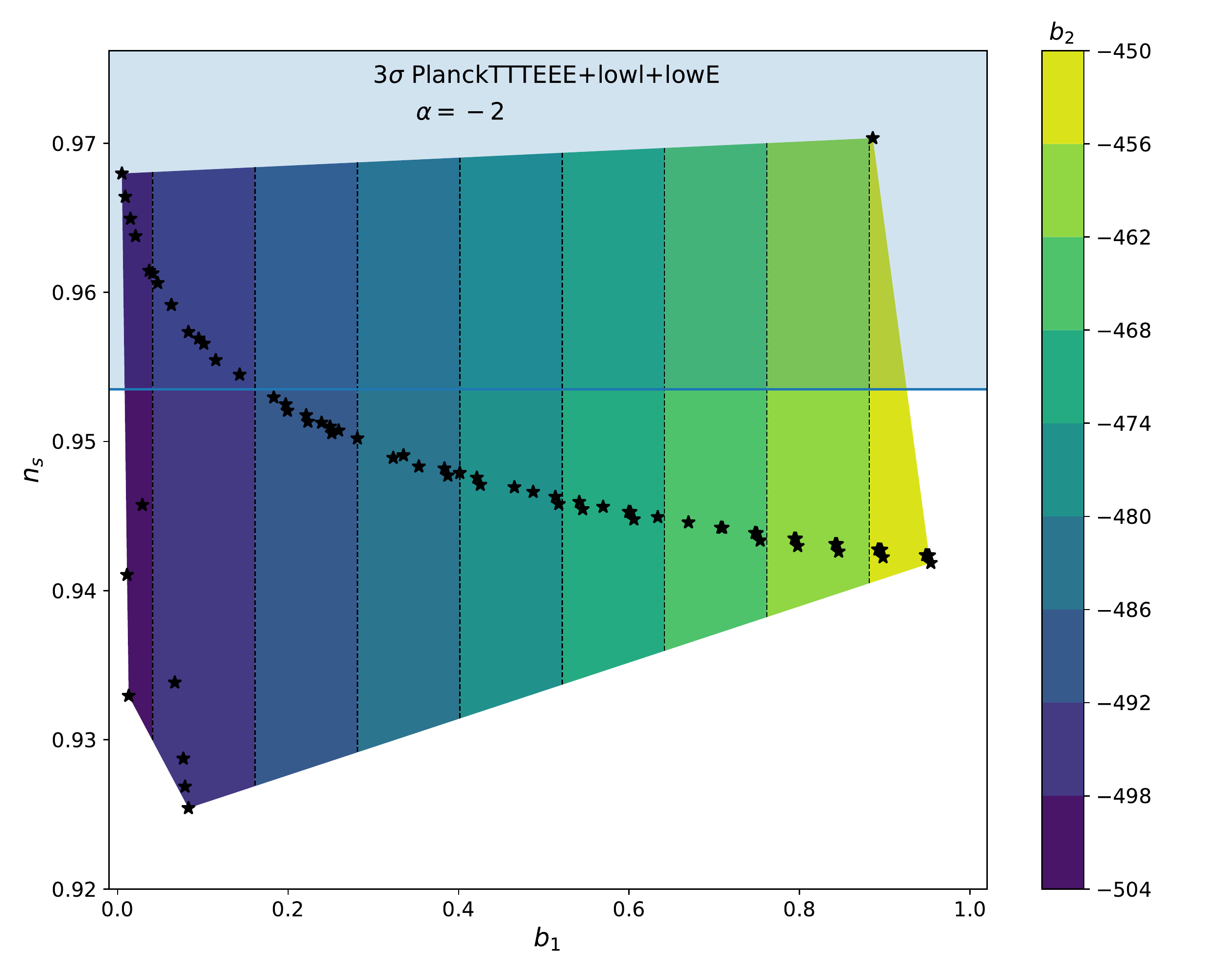}
        \includegraphics[width=0.45\columnwidth]{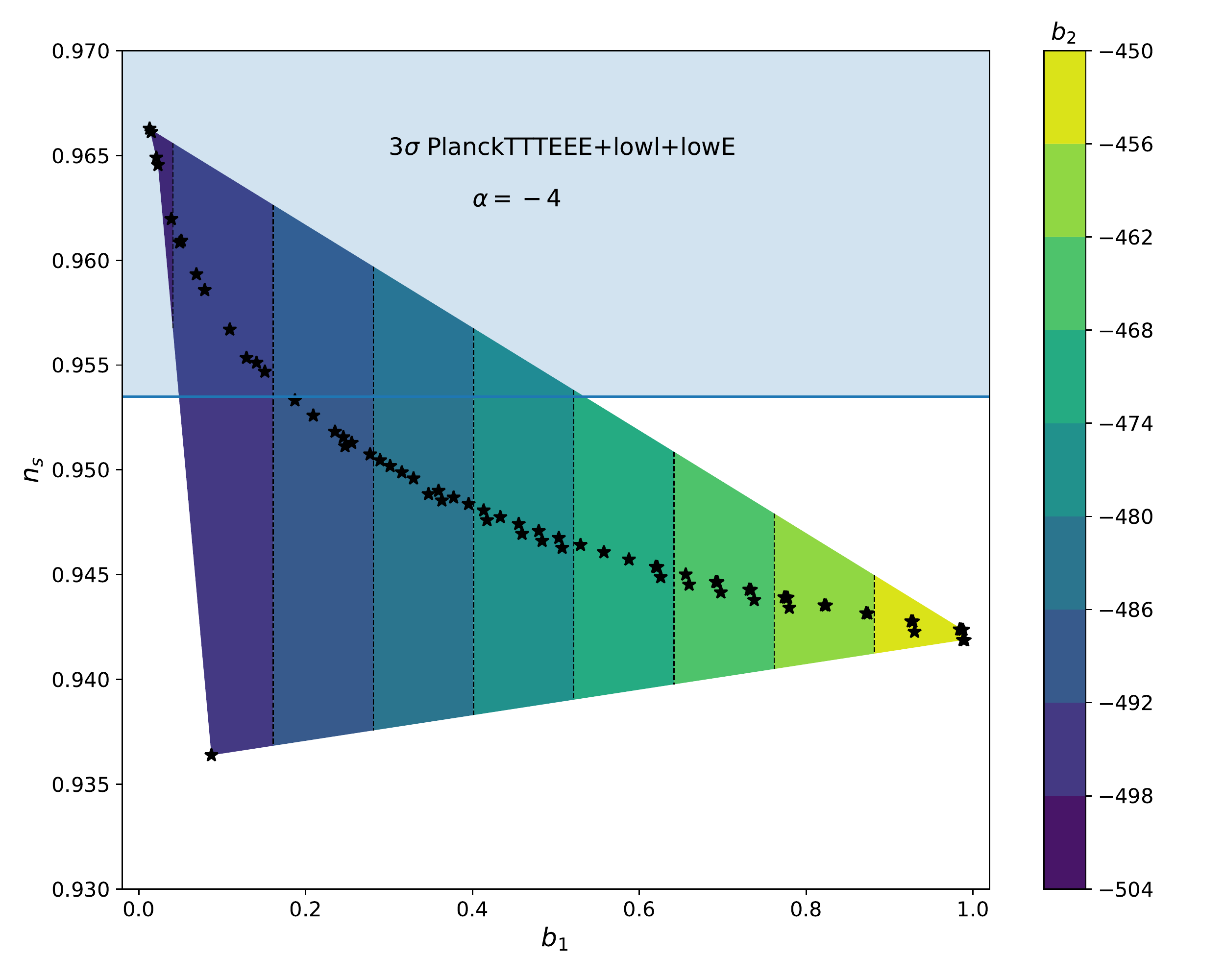}
        \includegraphics[width=0.45\columnwidth]{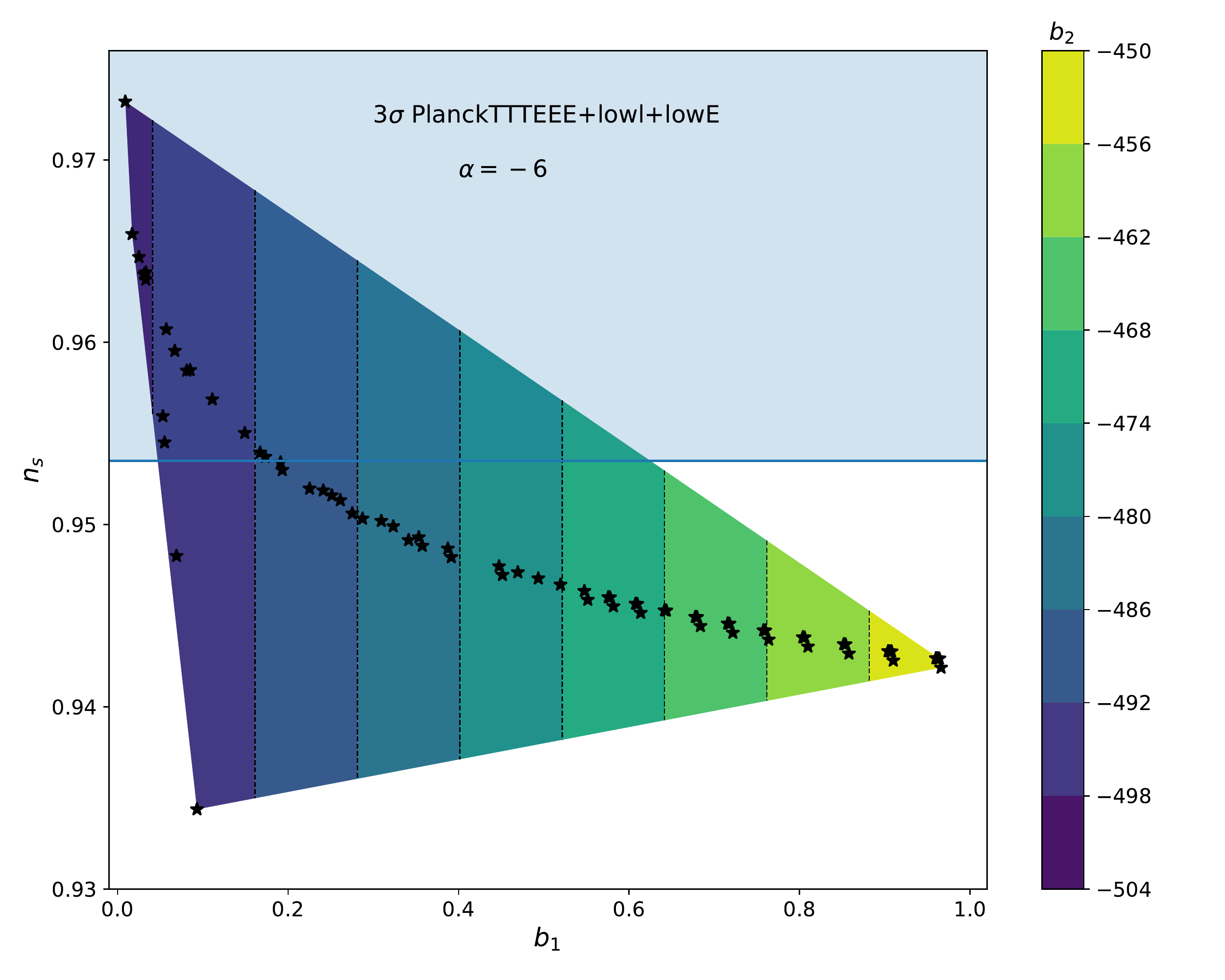} 
	\caption{Predictions for the scalar spectral index, $n_s$, for the model $U_1$. We consider different values for the model parameters $b_1$ and $b_2$, and fixed $\alpha$ and $c$, as indicated in each panel. The colored regions are the possible values for $n_s$ and the black points are the values that satisfy the slow-roll conditions. The light blue region stands for the $3\sigma$ C.L. of Planck temperature data, considering low multipoles and polarization data~\cite{Planck:2018}.}
	\label{fig:ns}
\end{figure}

\begin{figure}[ht!]
	\centering
	\includegraphics[width=0.45\columnwidth]{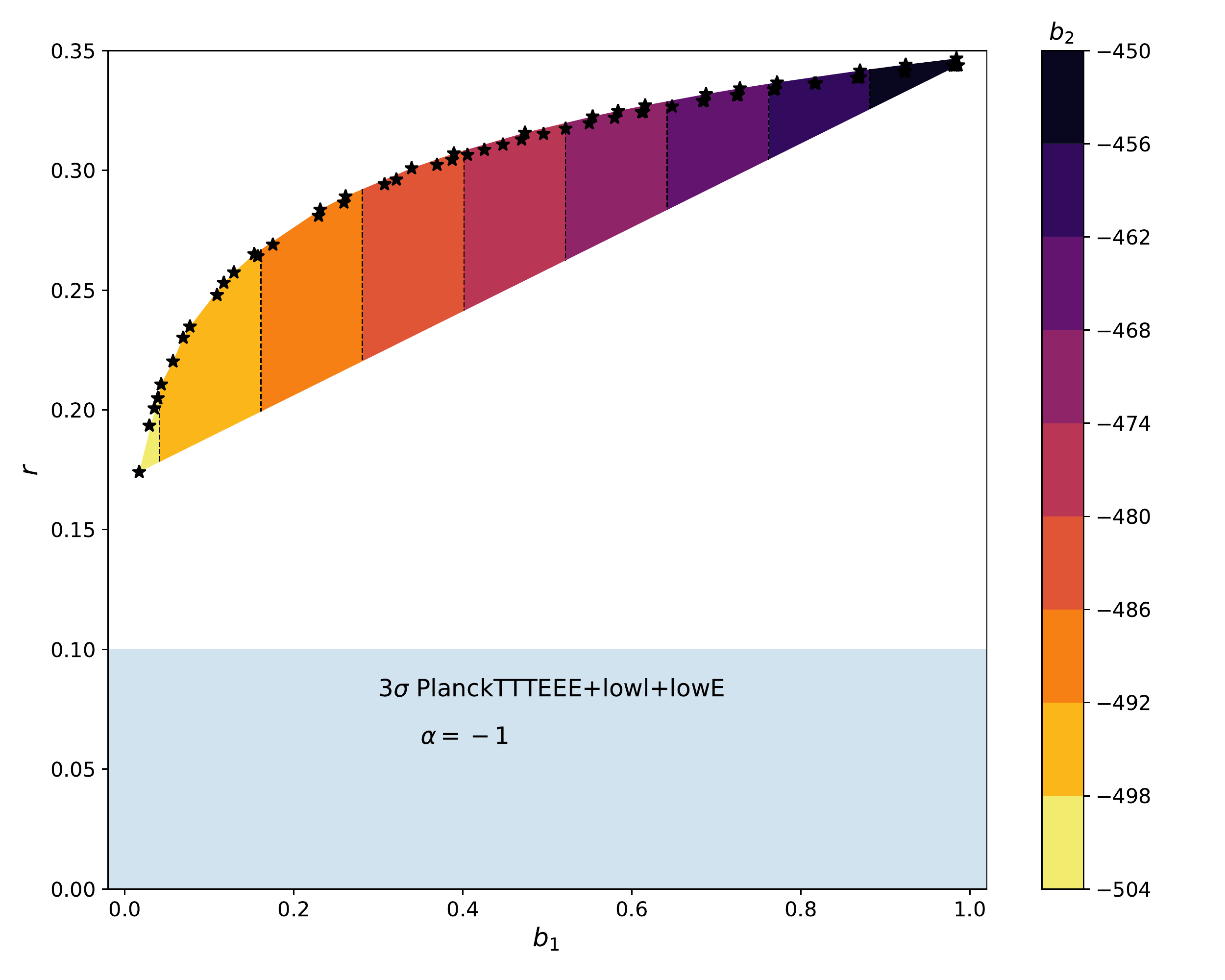}
        \includegraphics[width=0.45\columnwidth]{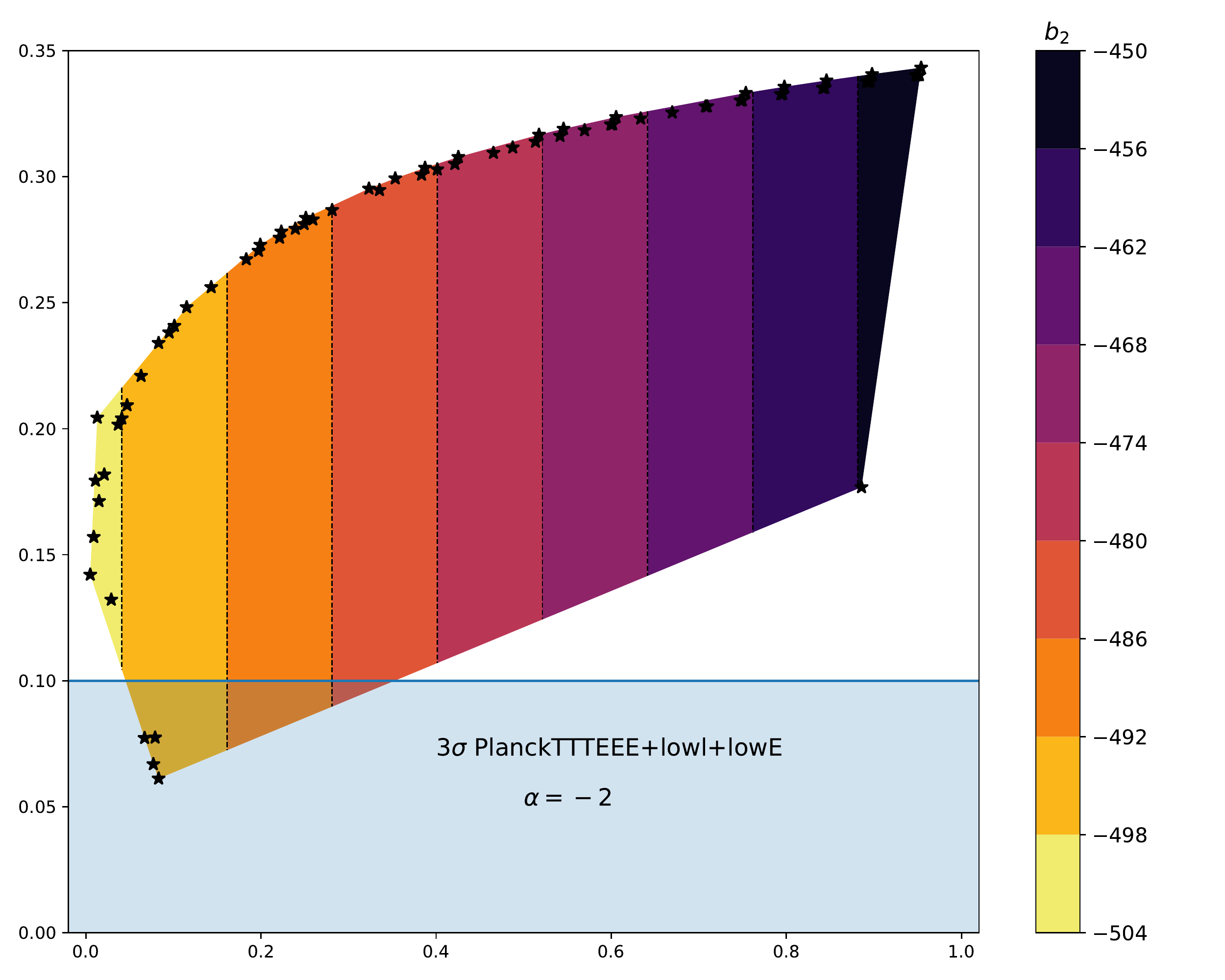}
        \includegraphics[width=0.45\columnwidth]{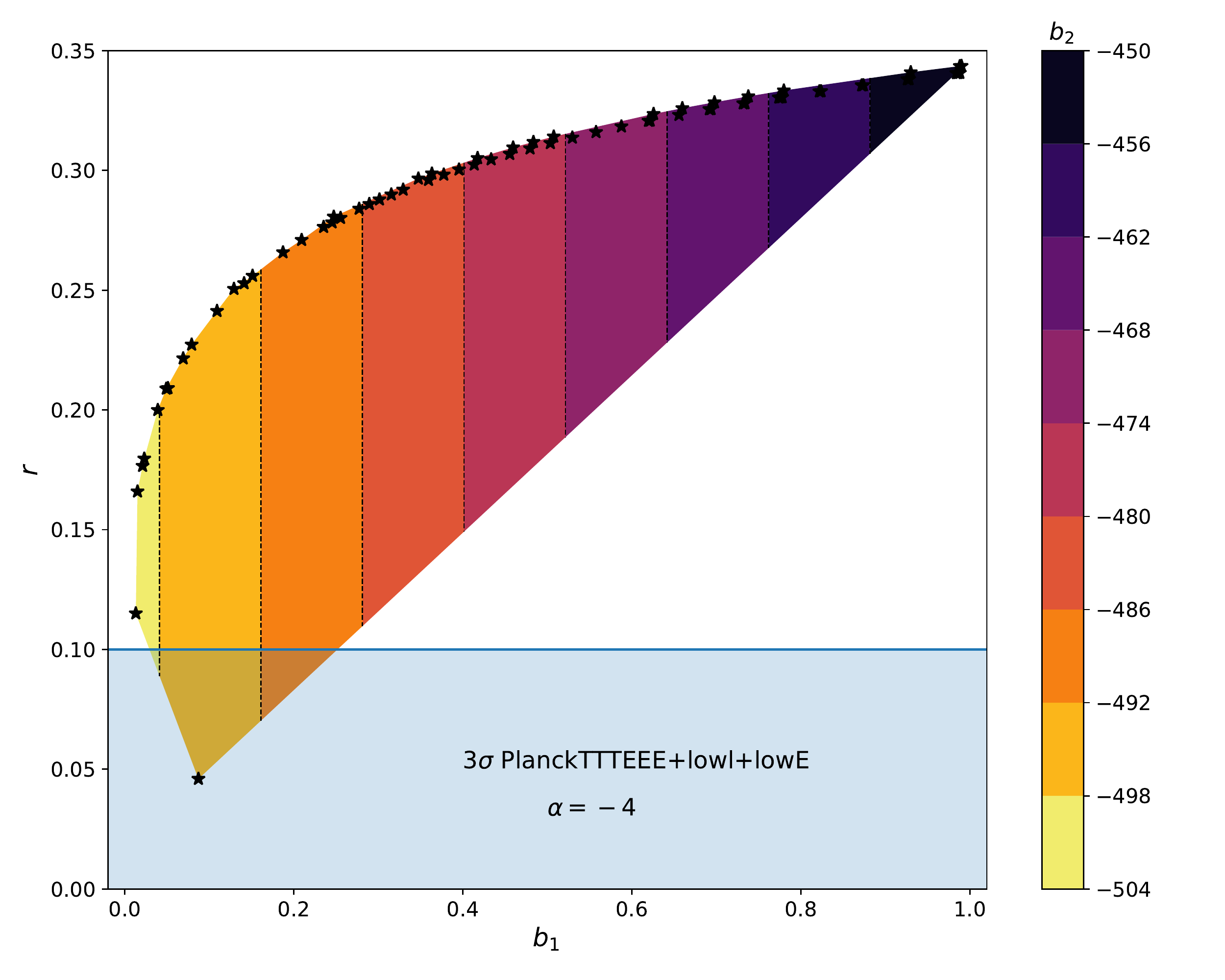}
        \includegraphics[width=0.45\columnwidth]{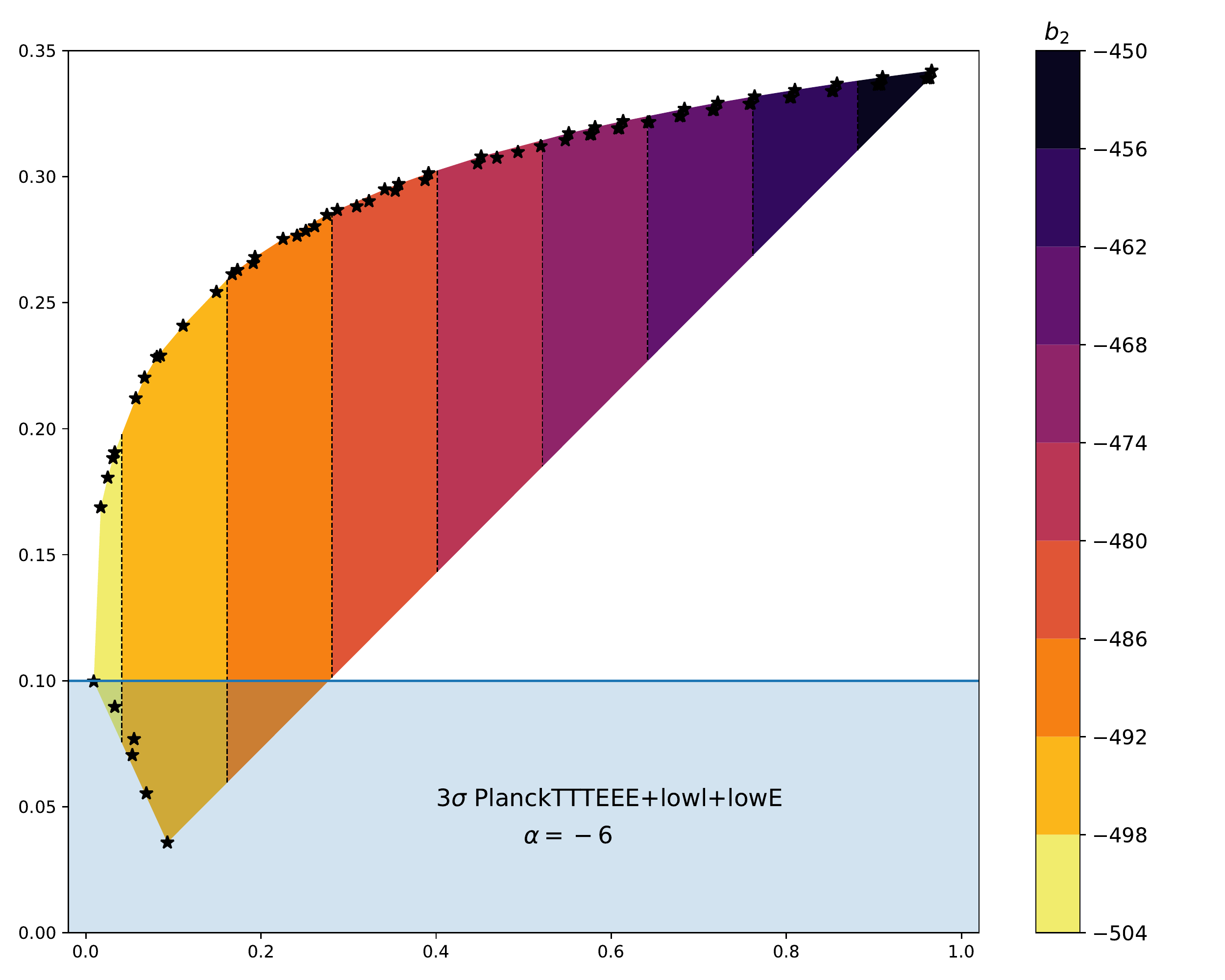} 
	\caption{Same as the Fig.\eqref{fig:ns}, but for the tensor-to-scalar ratio, $r$, for the model $U_1$.}
	\label{fig:r}
\end{figure}

In Figs.\eqref{fig:ns} and \eqref{fig:r} we show the behavior of $n_s$ and $r$, respectively, for the model $U_1$ with different values for the parameters $b_1,~ b_2,~ \alpha,~ c$, setting the number of e-folds $N=55$\footnote{By setting $N=55$ allow us to derive the value of the field when the CMB mode crosses the Hubble horizon during inflation.}. Note that we obtain a good agreement (at least at $3\sigma$ C.L. using the Planck data~\cite{Planck:2018}) for the scalar spectral index when $b_1<0.2$ and $-504<b_2<-492$, for all the $\alpha$ cases. Looking to the tensor-to-scalar ratio, the most promising cases happen meanwhile $-498<b_2<-492$, $b_1<0.2$, for $\alpha=-2$ and $\alpha=-4$, and $-504<b_2<-492$, $b_1<0.2$ for $\alpha=-6$. The case $\alpha=-1$ is completely discarded since it predicts $r>0.1$, which is excluded by CMB data. The fact that we obtain higher values for $r$, in this case, means that the energy scale for the inflaton field is too big to reproduce the CMB fluctuations when $\alpha=-1$.


However, the slow-roll conditions are not sufficient to put reasonable constraints on all these parameters, such that we need to extend the analysis to study the predictions for the CMB temperature power spectrum considering the full parameter space along with the matter density $\Omega_m$, the equation-of-state parameter $\omega$, etc. This kind of analysis is further justified since we obtain values of $n_s$ in agreement with Planck data, and, regarding this, we could investigate the impact of this class of models in the structure formation through the matter power spectrum. This analysis is currently under consideration and may appear in future work.

\section{Final Remarks}
\label{sec:04}

In this work, we show a new proposal for a theory of gravity, the $f(R, T) - \Lambda(\phi)$. We were able to derive its modified Friedmann equations and also the equation of motion for the inflaton field. By imposing a constraint that the inflaton field satisfies a first-order differential equation, we were able to derive analytic cosmological models. The methodology used to derive such models generalizes the results obtained by Moraes and Santos \cite{joaof}. Moreover, such a new methodology also enables us to find analytic cosmological models for non-trivial mappings between $f$ and $T$.

Both examples covered in this paper proved to be well-behaved in the radiation era, where $\omega = 1/3$ and $T = 0$. 
Besides, the extra terms coming from the running vacuum contributions can be used to fine-tune other cosmological parameters.
These good behaviors of the theory open a new path to test the viability of these models of gravity under cosmological data, such as those from the Planck Collaboration.
Moreover, the $f(R, T) - \Lambda(\phi)$ gravity stands for a proposal that rescues the inflationary eras driven by one single scalar field, generalizing the results presented in the beautiful paper of Ellis et al. \cite{Ellis:2014}.
Indeed, for the first model, $U_1$, we further explored the predictions for this primordial accelerated phase based on the equivalence between Jordan and Einstein frames. This approach allowed us to consider the slow-roll approximation and estimate the theoretical predictions for the scalar spectral index $n_s$ and the tensor-to-scalar ratio $r$. The results obtained present good agreement with Planck data, at least at $3\sigma$ C.L. for the parameter $n_s$. In addition, we can completely discard models with $\alpha<2$, since they predict higher values for $r$, in disagreement with CMB temperature data~\cite{Planck:2018}.

The methodology here presented can be extended to several other modified theories of gravity, such as in $f(Q)$ \cite{Jimenez:2017, Mandal:2020}, and $f(Q,T)$ \cite{Xu:2019, Arora:2020}. It also can be used to map non-trivial behavior of cosmological parameters in low redshift regimes which may appear in data coming from dedicated radio telescopes such as those from SKA \cite{Hall:2004} and from BINGO \cite{BINGO:01, BINGO:02}. Besides we can also add a term related to the presence of dust to verify its influence over the time evolution of our parameters \cite{Bazeia:2006_epjc}.  These discussions can be applied in braneworld scenarios as well \cite{Bazeia:2004}. We hope to be able to report on some of these topics in the near future.


{\acknowledgments} JRLS would like to thank CNPq (Grant nos. 420479/2018-0, and 309494/2021-4), and PRONEX/CNPq/FAPESQ-PB (Grant nos. 165/2018, and 0015/2019) for financial support. RSS would like to thank CAPES for financial support. SSC is supported by the Istituto Nazionale di Fisica Nucleare (INFN), sezione di Pisa, iniziativa specifica TASP.  The authors thank Jean Paulo Spinelli da Silva for reading the manuscript and offering suggestions and improvements. The authors are also grateful to the anonymous referee for comments and suggestions, which improved the quality of our work.

\end{document}